\documentclass{aa}
\usepackage[dvips]{graphicx}
\usepackage{psfrag}

\usepackage{natbib}

\def\lesssim{\mathrel{\hbox{\rlap{\hbox{\lower4pt\hbox{$\sim$}}}\hbox{$<$}}}}
\def\gtrsim{\mathrel{\hbox{\rlap{\hbox{\lower4pt\hbox{$\sim$}}}\hbox{$>$}}}}

\begin{document}

\title{Radio Emission from Cosmic Ray Air Showers:}

\subtitle{Coherent Geosynchrotron Radiation}

\author{T. Huege\inst{1} \and H. Falcke\inst{1,2,3}}


\institute{Max-Planck-Institut f\"ur Radioastronomie,
           Auf dem H\"ugel 69, 53121 Bonn, Germany\\
           email: thuege@mpifr-bonn.mpg.de and hfalcke@mpifr-bonn.mpg.de
         \and
           Radio Observatory, ASTRON, Dwingeloo, P.O. Box 2, 7990 AA Dwingeloo, The Netherlands
         \and
           Adjunct Professor, Dept. of Astronomy, University of Nijmegen, P.O. Box 9010, 6500 GL Nijmegen, The Netherlands
}

\date{Received April 2, 2003; accepted September 8, 2003}

\abstract{
Cosmic ray air showers have been known for over 30 years to emit pulsed radio emission in the frequency range from a few to a few hundred MHz, an effect that offers great opportunities for the study of extensive air showers with upcoming fully digital ``software radio telescopes'' such as LOFAR and the enhancement of particle detector arrays such as KASCADE Grande or the Pierre Auger Observatory. However, there are still a lot of open questions regarding the strength of the emission as well as the underlying emission mechanism. Accompanying the development of a LOFAR prototype station dedicated to the observation of radio emission from extensive air showers, LOPES, we therefore take a new approach to modeling the emission process, interpreting it as ``coherent geosynchrotron emission'' from electron-positron pairs gyrating in the earth's magnetic field.
We develop our model in a step-by-step procedure incorporating increasingly realistic shower geometries in order to disentangle the coherence effects arising from the different scales present in the air shower structure and assess their influence on the spectrum and radial dependence of the emitted radiation. We infer that the air shower ``pancake'' thickness directly limits the frequency range of the emitted radiation, while the radial dependence of the emission is mainly governed by the intrinsic beaming cone of the synchrotron radiation and the superposition of the emission over the air shower evolution as a whole. Our model succeeds in reproducing the qualitative trends in the emission spectrum and radial dependence that were observed in the past, and is consistent with the absolute level of the emission within the relatively large systematic errors in the experimental data.
}

\maketitle


\section{Introduction}

In the mid-1960s, \citet{JelleyFruinPorter1965} discovered that extensive air showers (EAS) initiated by high-energy cosmic rays produce strongly pulsed radio emission at frequencies around 40~MHz. The discovery triggered intensive research and in the following years a number of experiments established the presence of radio emission from EAS over a frequency-range from a few to a few hundred MHz. (For an excellent review of the historical developments and results we refer the interested reader to \citealt{Allan1971}.) Parallel to the experimental work, a number of authors worked on the theoretical interpretation of the emission processes (\citealt{KahnLerche1966}; \citealt{Lerche1967}; \citealt{Colgate1967}; \citealt{CastagnoliSilvestroPicchi1969} and \citealt{FujiNishimura1969}).

In the early 1970s, however, general interest started to focus on other, at the time more promising methods for air shower and cosmic ray research because of continuing technical difficulties involved in the radio measurements and problems with the interpretation of experimental data. Ground-based particle detectors and later on fluorescence techniques were so successful that activities concerning the radio frequency measurements of EAS virtually ceased. As a consequence, the research on radio emission from EAS froze on a rather basic level: while the empirical data gathered by the different experiments is largely discrepant, the theoretical models mentioned above adopt over-simplified geometries, do not incorporate relevant shower-characteristics such as realistic particle distributions or stay on a rather qualitative level that does not allow direct comparison with concrete experiments.

Today, over 30 years after the initial success of radio frequency measurements of EAS, the field is about to experience its renaissance. The availability of powerful digital data processing techniques and the advent of digital radio-interferometers such as LOFAR\footnote{http://www.lofar.org} (Low Frequency ARray) offer the realistic perspective to use radio frequency measurements of EAS as a very powerful and cost-effective tool that complements the established techniques very well. LOFAR, initially conceived for purely astronomical purposes, but offering incredible flexibility with its ability to form multiple simultaneous beams as made possible by its implementation in software, thereby builds a bridge between radio astronomy and particle physics.

Radio measurements of EAS share the main advantage of optical fluorescence techniques: They allow a very direct view into the development of the air shower and therefore yield information that profoundly simplifies the interpretation of data gained by ground based particle detectors. At the same time, however, they are not hindered by the need for superb observing conditions (clear, dark, moonless nights far away from any light pollution) that limits the duty cycle of optical fluorescence detectors to typically less than 10 \%. For a purely radio-triggered array with a low number of antennas, radio detection of EAS should be feasible for energies $\gtrsim 10^{17}$~eV. With large arrays such as LOFAR or in combination with external triggering by particle detector arrays such as KASCADE/KASCADE Grande \citep{AntoniApelBadea2003} or the Pierre Auger Observatory \citep{AugerCollaboration1996}, the study of EAS ranging from $\sim 10^{15}$~eV up to ultra-high energies would be possible \citep{FalckeGorham2003}. 

To investigate and develop the potential of LOFAR for radio frequency observations of EAS, we currently develop the LOfar Prototype Station LOPES \citep{HornefferFalckeKampert2002}, which is dedicated to the measurement of EAS. Obviously, its experimental realisation has to be accompanied by a thorough theoretical analysis of the underlying emission mechanism, since past theories have not been developed to sufficient depth for application to a concrete experiment such as LOPES.

In this work, we take a new approach to the theory of radio emission from EAS, namely the interpretation of the emission process as coherent synchrotron emission from electron-positron pairs deflected in the earth's magnetic field (or shorter: ``coherent geosynchrotron emission''), as proposed by \citet{FalckeGorham2003}; see also \citet{HuegeFalcke2002}. Other than \citet{SuprunGorhamRosner2003}, who recently simulated geosynchrotron emission from EAS with Monte Carlo techniques, we pursue an analytical approach to get a better understanding of the effects governing the emission.

We describe the basis of our approach in some detail in Sec.\ \ref{sec:geosynchrotron} and derive some observationally relevant quantities in Section 3. Sec.\ 4 summarises the characteristics of the air shower development that are needed for a realistic modeling of the emission process. In Sec.\ 5--8 we develop our model for the radio emission from EAS step by step with increasingly realistic geometries, which helps in understanding the coherence effects that play a role in shaping the emission spectrum and spatial distribution. After a short discussion of the results in Sec.\ 9 we conclude our work in Section 10.


\section{The geosynchrotron approach} \label{sec:geosynchrotron}

Two main emission mechanisms have been proposed in the past for radio emission from EAS: \v Cerenkov radiation from a charge excess moving with a velocity higher than the speed of light in the traversed medium (the so-called ``Askaryan'' mechanism motivated by \citealt{Askaryan1962a}; \citealt{Askaryan1965}) and acceleration of charged particles in the earth's magnetic field. While the former is dominant in case of dense media (\citealt{BuniyRalston2002}; \citealt{ZasHalzenStanev1992}; \citealt{AlvarezMunizVazquezZas2000}), polarisation measurements in a number of experiments subsequently supported the dominance of the geomagnetic emission mechanism for radio emission from EAS in air (e.g., \citealt{AllanClayJones1969}). It also seems unavoidable in principle for highly relativistic charged particles moving in the earth's magnetic field.

Coherent geosynchrotron emission from highly relativistic electron-positron pairs gyrating in the earth's magnetic field represents an equivalent scenario to that of the transverse currents of \citet{KahnLerche1966} (and other geomagnetic mechanisms) but is particularly appealing because it has the advantage of being based on well-studied and well-understood synchrotron theory, an excellent starting point for the development of our emission model. In the case of radio emission from cosmic ray air showers, however, coherence effects as well as non-periodic trajectories that are usually not considered for synchrotron radiation have to be taken into account.

In order to assess the coherence effects arising from the intrinsic air shower structure, we first analyse the emission from a specific point during the air shower evolution, namely the point of maximum shower development. Only in the last step we integrate over the shower evolution as a whole, which is effectively ``compressed'' into the radio pulse that the observer receives since the particles have velocities $v \approx c$.

At this stage, we do not take into account the Askaryan-type \v Cerenkov radiation. In other words, we set the refractive index of the atmosphere to unity.

\subsection{Synchrotron-theory: individual particles}

We base our calculations on the formalism developed in \citet{Jackson1975}. Any acceleration of a charge gives rise to electromagnetic radiation. The emission due to acceleration in the direction of the instantaneous velocity vector is, however, insignificant compared to that caused by the perpendicular acceleration \citep{Jackson1975}. As a consequence, any arbitrary particle motion, including the helical motion of a charged particle in a homogeneous magnetic field, can be approximated as an instantaneous circular trajectory with adequate curvature radius.

Retardation effects caused by the finite speed of light give rise to strong beaming effects for highly relativistic particles. For particles with Lorentz factor $\gamma$ the original dipole emission pattern is beamed into a narrow emission cone of order $\gamma^{-1}$ semi-opening angle which sweeps over the observer in a very short time, leading to strongly pulsed emission dominated by frequency components significantly higher than the particle gyration frequency.

   \begin{figure}
   \includegraphics[width=8.6cm]{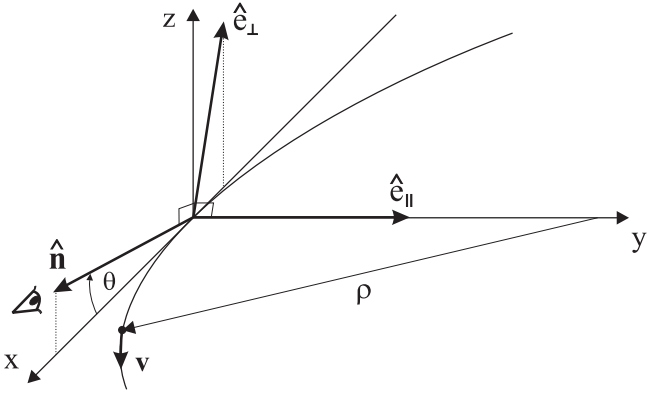}
   \caption{
   \label{fig:jackson_skizze}
   Geometry of single particle synchrotron radiation for an observer with line-of-sight vector $\vec{\hat{n}}$ enclosing a minimum angle $\theta$ to the instantaneous particle velocity vector $\vec{v}$. The equivalent curvature radius is given by $\rho$, and the emission can be conveniently divided into the components $\vec{\hat{e}}_{\perp}$ and $\vec{\hat{e}}_{\parallel}$. The particle trajectory lies in the x-y plane.
   }
   \end{figure}

The geometry of the problem corresponds to Fig.\ \ref{fig:jackson_skizze} if one chooses the origin of the coordinate system to lie in the point on the particle trajectory where the angle between instantaneous particle velocity vector $\vec{v}$ and line of sight vector $\vec{\hat{n}}$ reaches its minimum $\theta$.

Calculation in the frequency domain circumvents problems arising from the retardation effects. Jackson defines the quantity $\vec{A}(\vec{R},\omega)$ as a measure of the frequency component $\omega$ of the electric field normalised to unit solid angle $\Omega$. In the far-field limit (distance $R$ to the observer large compared to the extent of the particle trajectory, i.e.\ use of Fraunhofer-approximation is possible) $\vec{A}(\vec{R},\omega)$ can be approximated and conveniently divided into the two perpendicular components $\vec{\hat{e}}_{\perp}$ and $\vec{\hat{e}}_{\parallel}$ defined in Figure \ref{fig:jackson_skizze}. Retaining the phase information, $\vec{A}(\vec{R},\omega)$ can then be written as
\begin{equation}\label{eqn:singlestart}
\vec{A}(\vec{R},\omega) = \frac{\omega e}{\sqrt{8 c}\pi} \mathrm{e}^{\mathrm{i} (\omega \frac{R}{c}-\frac{\pi}{2})} \left[ -\vec{\hat{e}}_{\parallel} A_{\parallel}(\omega) \pm \vec{\hat{e}}_{\perp} A_{\perp}(\omega) \right],
\end{equation}
where the plus-sign is to be used for electrons and the minus-sign for positrons, $e$ denoting their unit charge. Furthermore
\begin{eqnarray}
A_{\parallel}(\omega) &=& \mathrm{i} \frac{2 \rho}{\sqrt{3} c} \left(\frac{1}{\gamma^{2}}+\theta^{2}\right) K_{2/3}(\xi),\\
A_{\perp}(\omega) &=& \theta \frac{2 \rho}{\sqrt{3} c} \left(\frac{1}{\gamma^{2}}+\theta^{2}\right)^{1/2} K_{1/3}(\xi)
\end{eqnarray}
with
\begin{equation}
\xi = \frac{\omega \rho}{3 c} \left(\frac{1}{\gamma^{2}}+\theta^{2}\right)^{3/2},
\end{equation}
where $\omega = 2 \pi \nu$ denotes the angular frequency corresponding to the observing frequency $\nu$, $K_{a}$ denotes the modified Bessel-function of order $a$, and the curvature radius of the instantaneous circular orbit is given by 
\begin{equation}\label{eqn:singleend}
\rho = \frac{v \gamma m_{e} c}{e B \sin \alpha}
\end{equation}
with magnetic field strength $B$ and pitch angle $\alpha$ between the particle trajectory and the magnetic field direction.

Apart from the adopted far-field approximations, the derivation of this result incorporates an integration over a highly oscillatory integrand only part of which contributes significantly. This integration is usually conducted using the so-called ``method of steepest descents'' also known as ``method of stationary phase'' \citep{Watson1944}. Jackson's derivation, although somewhat simplified, is correct as long as the observing frequency $\omega$ is high compared to the gyration frequency of the particles in the magnetic field. As the latter is around a few kHz and we are only interested in observing frequencies $>10$~MHz, the Jackson result is well suited as the basis for our calculations. It also correctly takes into account that the observer sees only one flash of radiation from each particle and not the periodic repetition that is associated with synchrotron radiation in the classical sense, since the mean free path length of the particles of $\sim 450$~m (at a height of 4~km) is very small compared with the length of a full gyration cycle of $\sim 20$~km.

The energy spectrum per unit solid angle of a single gyrating particle, correspondingly, is given by \citep{Jackson1975}
\begin{eqnarray}
\frac{\mathrm{d}^{2}I}{\mathrm{d}\omega \mathrm{d}\Omega} = 2 \left|\vec{A}(\vec{R},\omega)\right|^{2} & = & \frac{4e^{2}}{3\pi c^{2}}\left(\frac{\omega \rho}{c}\right)^{2}\!\left(\frac{1}{\gamma^{2}}+\theta^2\right)^{\!2}\\
& \times & \left[ K_{2/3}^{2}(\xi) + \frac{\theta^{2}}{\gamma^{-2}+\theta^{2}} K_{1/3}^{2}(\xi) \right].\nonumber
\end{eqnarray}
Since the energy spectrum is $\propto |\vec{A}(\vec{R},\omega)|^{2}$ it grows as $N^{2}$ with particle number $N$ if one assumes fully coherent emission. Given a specific distance to the observer $R$ the frequency component of the $E$-field can be calculated as
\begin{equation}\label{eqn:EofA}
\vec{E}(\vec{R},\omega) = \left(\frac{4 \pi}{c}\right)^{1/2} \frac{1}{R}\ \vec{A}(\vec{R},\omega).
\end{equation}
For a given (observer-frame) distribution of gyrating particles, the corresponding $\vec{E}(\vec{R},\omega)$ can then be superposed to calculate the total emission.

\subsection{Synchrotron-theory: electron-positron pairs}

   \begin{figure}
   \begin{center}
   \includegraphics[height=5.3cm]{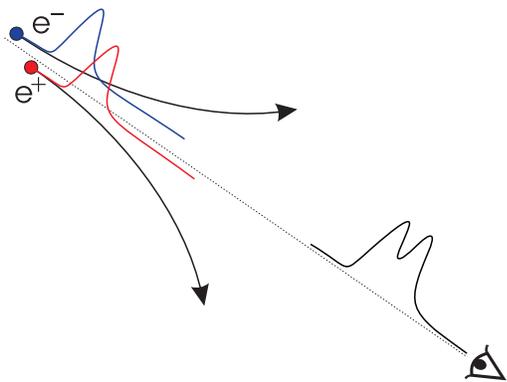}
   \caption{
   \label{fig:paircoherence}
   Misalignment between the electron and the positron in an electron-positron pair no longer allows coherent addition of the individual emissions.
   }
   \end{center}
   \end{figure}

In the air shower, electrons and positrons are created in pairs. The symmetry arising from the opposite curvature of electron and positron trajectories can lead to a significant simplification of the calculation: For an electron-positron pair with {\em perfectly} symmetric trajectories with regard to the observer, the $A_{\parallel}$ contributions from the two particles add up to $2 A_{\parallel}$, whereas the $A_{\perp}$ contributions completely cancel each other.

This is, however, an overly special case which does not adequately represent the problem we are facing. Depending on the direction from which the observer sees the particle pair, the cancellation of the $A_{\perp}$ contributions as well as the summation of the $A_{\parallel}$ contributions are only partial. Furthermore, as the pulses emitted by the relativistic particles are very short, there is an inherent coherence length associated to the emissions of the individual particles. If there is considerable misalignment between the particles, the resulting phase differences destroy the coherence as illustrated in Figure \ref{fig:paircoherence}. Overall, one would therefore have to quantify the coherence losses and incomplete summation/cancellation arising from the pairing through a form factor.

A more detailed look at the numbers and characteristics of the particle distributions in the shower, however, reveals that we may indeed assume full summation and cancellation of $A_{\parallel}$ and $A_{\perp}$ for an ``effective'' electron-positron pair without introducing too large an error. This approximation works well if we no longer look at electron-positron pairs that actually form together but rather group pairs of positrons and electrons together such that their trajectories overlap symmetrically as seen by the observer --- and if we can accomplish this pairing for the vast majority of particles.

For coherent addition of the positron and electron emission to be possible, a significant portion of those parts of the particle trajectories from which the observer actually receives radiation must overlap. (That part has a length of $\sim 110$~m for $\gamma=60$, given by the length over which the instantaneous velocity vector encloses an angle $\lesssim \gamma$ with the observer's line of sight.) In a typical $10^{17}$~eV air shower the shower ``pancake'', even somewhat before and after the shower's maximum development, consists of $\sim 10^{8}$ particles at any time. Even if the particles were distributed homogeneously in the shower pancake, this would lead to a particle density of $\sim 1000$~m$^{-3}$. For the realistic distributions described in section \ref{sec:shower}, the densities in the dominating centre region are a lot higher. This illustrates that each particle (except in the unimportant outskirts of the shower pancake) will a priori have a high number of particles in its direct vicinity with which it can be paired. The probability that there is a significant overlap between the paired particles' trajectories is high because the particles' mean free path length of $\sim 450$~m is considerably larger than the aforementioned $\sim 110$~m of the trajectory from which the observer receives radiation.

Whether a consequent pairing with {\em symmetric trajectories} is possible, however, depends critically on the direction distribution of the particles' instantaneous velocity vectors. Throughout this work we assume a $\delta$-distribution of the particle velocity directions at any given point in the shower shell, as we choose the initial velocity vectors to point radially away from the spherical shower surface. In this situation, the pairing of particles with symmetric trajectories becomes simple as long as one allows pairing between positrons and electrons from generations of particles with a certain net offset in generation time.

   \begin{figure}
   \psfrag{theta01ogamma}[c][b]{$\theta\ [\gamma^{-1}]$}
   \includegraphics[width=8.6cm]{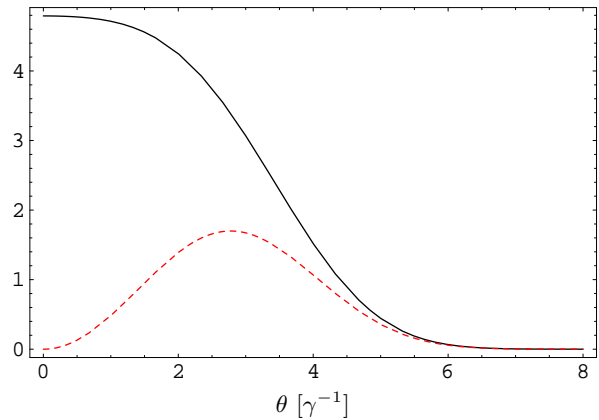}
   \caption{Comparison of $|A_{\parallel}|$ (solid) and $|A_{\perp}|$ (short-dashed) for $\nu=100$~MHz, $\gamma=60$ and $B=0.3$~G. Absolute scale is arbitrary.
   \label{fig:apvsasplot}}
   \end{figure}

In this scenario of high particle density and $\delta$-distribution of velocity directions, the emission from an ``effective'' electron-positron pair can therefore be approximated as that from a pair with perfectly symmetric trajectories:
\begin{equation}\label{eqn:Epair}
\vec{E}_{\mathrm{p}}(\vec{R},\omega) \approx \left(\frac{4 \pi}{c}\right)^{1/2}\! \frac{1}{R}\ \frac{2 \omega e}{\sqrt{8 c}\pi} \mathrm{e}^{\mathrm{i} (\omega \frac{R}{c}-\frac{\pi}{2})}  \left(-\vec{\hat{e}}_{\parallel}\right) A_{\parallel}(\omega).
\end{equation}
The fact that $|A_{\parallel}| > |A_{\perp}|$, especially for small $\theta$ where most of the radiation is emitted (Fig.\ \ref{fig:apvsasplot}), furthermore demonstrates that $A_{\perp}$ is not dominating the emission, anyway, and therefore gives further confidence in the approximation.

Effectively, this result allows us to drop the differentiation between positrons and electrons and to consider only generic ``particles'' hereafter. The spectrum emitted by such an individual particle then corresponds to:
\begin{equation}\label{eqn:Epairpart}
\vec{E}(\vec{R},\omega) = \left(\frac{4 \pi}{c}\right)^{1/2} \frac{1}{R}\ \frac{\omega e}{\sqrt{8 c}\pi} \mathrm{e}^{\mathrm{i} (\omega \frac{R}{c}-\frac{\pi}{2})}  \left(-\vec{\hat{e}}_{\parallel}\right) A_{\parallel}(\omega).
\end{equation}
Superposition of these spectra for all particles in the shower, correctly taking into account the phase differences arising from their relative positions, then yields the emission from the air shower as a whole.

\section{Observational quantities}

We present a number of relations of the previous results to observational quantities:

\subsection{Pulse reconstruction}

The time-dependence of the electromagnetic pulse corresponding to a given spectrum $\vec{E}(\vec{R},\omega)$ can easily be reconstructed for a specific receiver bandwidth by an inverse Fourier-transform of the remaining spectrum. Hence, if the frequency characteristic of the receiver is given by $b(\omega)$, the time-dependence of the electric field $\vec{E}(\vec{R},t)$ can be calculated as
\begin{equation}\label{eqn:pulserec}
\vec{E}(\vec{R},t)=\frac{1}{\sqrt{2 \pi}} \int_{-\infty}^{\infty} b(\omega) \vec{E}(\vec{R},\omega)\ \mathrm{e}^{-\mathrm{i}\omega t}\ \mathrm{d}\omega,
\end{equation}
where $\vec{E}(\vec{R},-\omega)$ is given by the complex conjugate of $\vec{E}(\vec{R},\omega)$.

\subsection{Conversion of $\left|\vec{E}(\vec{R},\omega)\right|$ to $\epsilon_{\nu}$}

In the works of the 1960ies and 1970ies, the strength of the measured radio emission was usually denoted with a quantity $\epsilon_{\nu}$ in units of $\mu$V~m$^{-1}$~MHz$^{-1}$, which was defined as the peak electric field strength during the pulse divided by the effective bandwidth of the receiver system $\Delta\nu$. In practice, the total pulse amplitude (in V) at a given observing frequency $\nu$ was derived from the photographed oscilloscope traces of the two polarisation directions and then converted to an electric field strength (in V/m) taking into account the receiver and antenna gain. This field strength, representing the projection of the electric field vector on the horizontal plane, was then ``back-projected'' to yield the field strength in the plane perpendicular to the shower axis and the magnetic field, in which the electric field vector lies (see Eq.\ (\ref{eqn:epardir}) for $\vartheta=0$). Division of the resulting field strength by the effective bandwidth $\Delta\nu$ of the receiver system then yielded $\epsilon_{\nu}$.

To compare our theoretical values of $\left|\vec{E}(\vec{R},\omega)\right|$ to the experimental results for $\epsilon_{\nu}$, we analytically reconstruct the time-dependence of the electromagnetic field pulse $\vec{E}(t)$ for the case of an idealised rectangle filter of bandwidth $\Delta\nu$, over which $\left|\vec{E}(\vec{R},\omega)\right|$ is assumed to be constant, and which is centred on the observing frequency $\nu$. After time-averaging over the high-frequency oscillations, $\epsilon_{\nu}$ then directly follows from the peak field amplitude divided by $\Delta\nu$ and is given by
\begin{equation} \label{eqn:econversion}
\epsilon_{\nu}=\sqrt{\frac{128}{\pi}}\left|\vec{E}(\vec{R},\omega)\right| \approx 6.4 \left|\vec{E}(\vec{R},\omega)\right|.
\end{equation}

\subsection{LOPES signal-to-noise} \label{sec:lopessnr}

In order to compare our predictions to the abilities of LOPES (or any other generic dipole array), we first estimate the expected signal-to-noise ratio (SNR) for a receiving system consisting of an individual inverted V shape dipole antenna with gain $G=1.9$ and a receiver incorporating a filter with bandwidth $\Delta\nu$ centred on the observing frequency $\nu$, a square-law detector (i.e., a detector measuring the received power) and an integrator which averages the signal over a time $\tau$.

The noise level of the receiving system, in our case dominated by Galactic noise, can be characterised by the noise temperature $T_{\mathrm{sys}}\approx T_{\mathrm{sky}}(\nu)$. Comparison with the temperature increase $\Delta T$ corresponding to the power of the pulse intercepted by the antenna then yields
\begin{equation} \label{eqn:snr}
\mathrm{SNR}=\sqrt{2\,\Delta\nu\,\tau}\ \frac{\Delta T}{T_{\mathrm{sys}}},
\end{equation}
where the first factor takes into account the increase of the SNR due to the number of independent samples accumulated in case of bandwidth $\Delta\nu$ and averaging time $\tau$ as determined by the Nyquist theorem. The energy flux of an electromagnetic wave propagating through the (vacuum-like) atmosphere is given by the Poynting vector, {\em{in SI-units}} and omitting the argument $\vec{R}$ for the fields being defined as
\begin{equation}
\vec{S}(t)=\vec{E}(t)\times\vec{H}(t)=\frac{1}{\mu_{0}}\vec{E}(t)\times\vec{B}(t),
\end{equation}
where $\mu_{0}=4\pi\ 10^{-7}$~Vs/Am. As $\vec{E} \perp \vec{B}$, it follows that
\begin{equation}
\left|\vec{S}(t)\right|=\frac{1}{c \mu_{0}}\left|\vec{E}(t)\right|^{2}\approx\frac{1}{120\pi\,\Omega}\left|\vec{E}(t)\right|^{2}.
\end{equation}
For a point-like source, the effective area of a dipole antenna is given by \citep{RohlfsWilson1996} $A_{\mathrm{eff}}=G \lambda^{2}/4\pi=G c^{2}/4\pi\nu^{2}$, so that it receives the power
\begin{equation}
P(t)=\frac{1}{2}A_{\mathrm{eff}} \left|\vec{S}(t)\right|=\frac{G c}{8\pi\nu^{2}\mu_{0}} \left|\vec{E}(t)\right|^{2},
\end{equation}
where the factor 1/2 is introduced for an antenna measuring only one polarisation direction of unpolarised radiation. Averaging over the time $\tau$ then leads to
\begin{eqnarray}
<\!P\!>_{\tau}&=&\frac{G c}{8\pi\nu^{2}\mu_{0}\tau} \int_{0}^{\tau}\!\left|\vec{E}(t)\right|^{2}\ \mathrm{d}t\nonumber\\
&\approx&\frac{G c}{8\pi\nu^{2}\mu_{0}\tau} \int_{\omega_{l}}^{\omega_{h}}\!\left|\vec{E}(\omega')\right|^{2}\ \mathrm{d}\omega',
\end{eqnarray}
where $\omega_{h/l}=2\pi(\nu \pm 1/2\Delta\nu)$ and the last step follows from Parseval's theorem as long as the bulk of the pulse is sampled in the averaging time $\tau$. Assuming that the spectrum is flat over the observing bandwidth $\Delta\nu$ with a value $\left|\vec{E}(\omega')\right|\equiv\left|\vec{E}(2\pi\nu)\right|=\mathrm{const.}$ and using Eq.\ (\ref{eqn:econversion}), we can write
\begin{eqnarray}
<\!P\!>_{\tau}&\approx&\frac{G c}{8\pi\nu^{2}\mu_{0}\tau} \left|\vec{E}(2\pi\nu)\right|^{2} 2\pi\Delta{\nu}\nonumber\\
&=&\frac{G c}{4\nu^{2}\mu_{0}\tau} \frac{\pi}{128}\epsilon_{\nu}^{2} \Delta\nu.
\end{eqnarray}
This averaged power is then directly related to $\Delta T$ via the Boltzmann-constant $k_{\mathrm{B}}$ by
\begin{equation}
\Delta T = \frac{<\!P\!>_{\tau}}{k_{\mathrm{B}} \Delta\nu},
\end{equation}
so that from Eq.\ (\ref{eqn:snr}) follows
\begin{equation}
\mathrm{SNR} \approx \frac{\pi\,G c}{256\,\sqrt{2}\,\nu^{2}\mu_{0}k_{\mathrm{B}}}  \sqrt{\frac{\Delta\nu}{\tau}} \frac{\left|\epsilon_{\nu}\right|^{2}}{T_{\mathrm{sys}}}.
\end{equation}
Setting $\tau$ to a sensible value of $2\Delta\nu^{-1}$, we get
\begin{eqnarray}
\mathrm{SNR} &\approx& 0.5 \left(\frac{G}{1.9}\right) \left(\frac{\nu}{60\ \mathrm{MHz}}\right)^{-2} \left(\frac{T_{\mathrm{sky}}(\nu)}{4000\ \mathrm{K}}\right)^{-1} \nonumber\\
&\times& \left(\frac{\Delta\nu}{35\ \mathrm{MHz}}\right) \left(\frac{\epsilon_{\nu}}{1\ \mu\mathrm{V}\ \mathrm{m}^{-1}\ \mathrm{MHz}^{-1}}\right)^2 
\end{eqnarray}
for a typical LOPES antenna with an observing bandwidth of 35~MHz centred on the observing frequency 60~MHz and an estimate of $T_{\mathrm{sky}}(60\ \mathrm{MHz}) \approx 4,000$~K \citep{FalckeGorham2003}.

For a complete LOPES array consisting of $N_{\mathrm{ant}}$ antennas, the SNR is then increased by an additional factor $\sqrt{1/2N_{\mathrm{ant}}(N_{\mathrm{ant}}-1)}\approx \sqrt{1/2} N_{\mathrm{ant}}$ for large $N_{\mathrm{ant}}$.

\section{Extensive air shower properties}\label{sec:shower}

Extensive air showers can be initiated by primary particles with strongly differing energy and composition and at variable inclination angles. Consequently, their properties such as the position of their maximum development and the longitudinal and lateral distributions of secondary particles can vary significantly.

Additionally, the simulation of air showers consisting of $> 10^{8}$ particles with energies in the MeV range created in a cascade initiated by primary particles of energies as high as $10^{20}$ eV, is in itself a very difficult process. There are elaborate Monte Carlo simulations such as CORSIKA \citep{HeckKnappCapdevielle1998} and AIRES \citep{Sciutto1999} which themselves incorporate a number of different models for the underlying particle interactions. But although these codes are very sophisticated, uncertainties remain, especially at the very highest energies that are out of the reach of accelerator experiments (see, e.g., \citealt{KnappHeckSciutto2003}).

At this stage, however, we do not incorporate the results of elaborate air shower simulations. We rather revert to the widely used analytical parametrisations for the longitudinal development and lateral particle distributions dating back to \citet{Greisen1956}, \citet{KamataNishimura1958} and \citet{Greisen1960} which are admittedly crude, but as a first step seem adequate to describe the properties relevant to our calculations that an ``average'' air shower would have in case of vertical inclination. (For an overview see, e.g., \citealt{Gaisser1990}.)

\subsection{Longitudinal air shower development}\label{sec:shower:longdev}

The longitudinal air shower development can be parametrised by the so-called ``shower age'' $s$ as a function of atmospheric depth $X$:
\begin{equation}
s(X)=\frac{3X/X_{0}}{(X/X_{0})+2 \ln(E_{\mathrm{p}}/E_{\mathrm{crit}})}=\frac{3X}{X+2X_{\mathrm{m}}}
\end{equation}
where $X_{0}=36.7$~g~cm$^{-2}$ denotes the electron ``radiation length'' in air, $E_{\mathrm{crit}}=86$~MeV corresponds to the threshold energy where ionisation losses equal radiation losses for electrons moving in air, and $X_{\mathrm{m}}=X_{0} \ln(E_{\mathrm{p}}/E_{\mathrm{crit}})$ marks the theoretical value for the depth of the shower maximum in this parametrisation. The shower commences at $s=0$, builds to its maximum development at $s=1$ and then declines over the range $s=1$--3. Although originally developed for purely electromagnetic showers, the formula is suitable to qualitatively describe the average development of the ``clumpier'' hadronic showers as well. The theoretical $X_{\mathrm{m}}$ value does then, however, not correspond to the actual position of the shower maximum. For purely electromagnetic showers, the development of the total number of charged particles (almost purely electrons and positrons) can be described by
\begin{equation}\label{eqn:Nelongitudinal}
N(s)=\frac{0.31 \exp\left[(X/X_{0})(1-\frac{3}{2} \ln s)\right]}{\sqrt{\ln(E_{\mathrm{p}}/E_{\mathrm{crit}}})}
\end{equation}
as a function of shower age. The predicted value for $N$ in the shower maximum ($s=1$) is very close to the \citet{Allan1971} ``rule of thumb'' $N_{\mathrm{max}}=E_{\mathrm{p}}/\mathrm{GeV}=10^{8}$ for a 10$^{17}$~eV shower. For the position of the shower maximum $X_{\mathrm{max}}$ we refer to the measurements and simulations presented in \citet{KnappHeckSciutto2003} that suggest a value of $X_{\mathrm{max}}\approx 630$~g~cm$^{-2}$ which corresponds to $R_{0}\approx 4$~km for a $10^{17}$~eV air shower and to the works of \citet{Pryke2001} as well as \citet{AbuZayyadBelovBird2001}.

\subsection{Lateral particle distribution}\label{sec:lateraldistri}

The lateral particle density can be described with the NKG (Nishimura-Kamata-Greisen) parametrisation, which without normalisation corresponds to

\begin{eqnarray}
\rho_{\mathrm{NKG}}(r) &=& \frac{1}{r_{\mathrm{M}}^{2}}\ \frac{\Gamma(4.5-s)}{2\pi\Gamma(s)\Gamma(4.5-2s)}\nonumber\\ &\times&\left(\frac{r}{r_{\mathrm{M}}}\right)^{s-2} \left(1+\frac{r}{r_{\mathrm{M}}}\right)^{s-4.5}.
\end{eqnarray}
To avoid the unphysical singularity of the NKG profile at the shower centre we cut off the distribution with a constant value at radii smaller than 0.1~m. The normalisation for the integration is chosen correspondingly (see Sec.\ \ref{sec:integration}).

The parameters relevant to the NKG distribution, shower age $s$ and Moli\`ere radius $r_{\mathrm{M}}$, show a high degree of degeneracy. The increase in $s$ during the shower propagation broadens the lateral distribution, but at the same time the decrease of $r_{\mathrm{M}}$ with increasing atmospheric density tends to narrow it. One can therefore often parametrise a given lateral particle distribution with a wide range of different values for $s$ and $r_{\mathrm{M}}$, where $r_{\mathrm{M}}$ in fact need not be close to the theoretical Moli\`ere radius at all \citep{AntoniApelBadea2001}. We here stick to the theoretically motivated values of $s=1$ for the shower maximum and set $r_{\mathrm{M}}$ to the Moli\`ere radius at the atmospheric height of the maximum derived from the atmospheric density as \citep{DovaEpeleMariazzi2003}
\begin{equation}
r_{\mathrm{M}}(h)=r_{\mathrm{M}}(h_{0}) \frac{\rho_{\mathrm{atm}}(h_{0})}{\rho_{\mathrm{atm}}(h)}=9.6\ \frac{\mathrm{g\ cm}^{-2}}{\rho_{\mathrm{atm}}(h)}.
\end{equation}
According to the US standard atmosphere of 1977 as implemented in CORSIKA \citep{Ulrich1997} the atmospheric density at a height of 4~km corresponds to $\rho_{\mathrm{atm}}=0.82$~mg~cm$^{-3}$, which in turn yields a Moli\`ere radius of $r_{\mathrm{M}}\approx 117$~m.

\subsection{Particle arrival time distribution}\label{sec:shower:arrivalt}

Knowledge of the arrival time distributions of particles in the air shower is necessary to parametrise the curvature and thickness of the shower front as a function of radial distance to the core. Unfortunately, the development of the particle arrival time distributions during the shower evolution is not well established. \citet{AgnettaAmbrosioAramo1997} have analysed Haverah Park data of more than 450,000 air shower events. These lie in the adequate energy range of $\sim 10^{17}$~eV, but were measured at an altitude of 220~m and cannot differentiate between e$^{\pm}$ and $\mu^{\pm}$. They, however, still trace the distribution of e$^{\pm}$ correctly because the number of e$^{\pm}$ by far exceeds the number of $\mu^{\pm}$.  An earlier analysis of Volcano Ranch data by \citet{Linsley1986} reflects the particle distribution at an altitude of 1,800~m, but is based on a very low number of events (especially at the radial distances up to a few hundred metres relevant to our model) and only determines the shower thickness and not the functional form of the arrival time distributions. We therefore base our model on the \citet{AgnettaAmbrosioAramo1997} data and use the \citet{Linsley1986} data only for comparison. 

\begin{figure}
   \psfrag{t0ns}[c][t]{$t$~[ns]}
   \includegraphics[width=8.6cm]{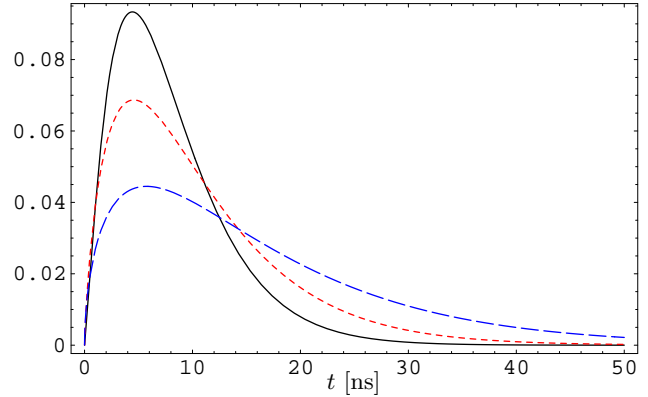}
   \caption{$\Gamma$-pdf determining the arrival time distribution of particles as measured by \protect\citet{AgnettaAmbrosioAramo1997}. Solid: in the shower centre, short-dashed: 50~m from shower centre, long-dashed: 100~m from shower centre.
   \label{fig:gammapdf_plot}}
   \end{figure}

In \citet{AgnettaAmbrosioAramo1997} the measured arrival time distribution at a given radial distance is fitted with a $\Gamma$-probability distribution function ($\Gamma$-pdf) defined as
\begin{equation}\label{eqn:gampdf}
f(t)=A\ t^{B}\ \exp(-Ct),
\end{equation}
the form of which (cf.\ Fig. \ref{fig:gammapdf_plot}) arises from multiple scattering events during the shower propagation. While $A$ only comprises a normalisation factor, the fit parameters $B$ and $C$ of the $\Gamma$-pdf are directly related to the mean arrival time $<\!t\!>$ and the standard deviation $\sigma_{t}$ of the measured arrival time distributions \citep{Bury1975},
%
\begin{equation}
B=\left(\frac{<\!t\!>}{\sigma_{t}}\right)^{2}-1 \quad\mathrm{and}\quad C=\frac{<\!t\!>}{\sigma_{t}^{2}}.
\end{equation}
The dependence of $<\!t\!>$ and $\sigma_{t}$ on the radial distance to the shower core is modeled by a generalised paraboloid of the form
\begin{equation}
<\!t\!>\!(r)\ \!,\ \!\sigma_{t}(r) = F +\ \!G\ \left(r/r_{0}\right)^{H}
\end{equation}
where $r_{0}$ is set to the Moli\`ere radius at ground level of 79~m. The parameter sets for $<\!t\!>\!(r)$ and $\sigma_{t}(r)$ are listed as
\begin{eqnarray}
F_{t} &=& (8.039 \pm 0.068)\ \mathrm{ns}\nonumber\\
G_{t} &=& (5.508 \pm 0.095)\ \mathrm{ns}\\
H_{t} &=& 1.710 \pm 0.059\nonumber
\end{eqnarray}
and
\begin{eqnarray}
F_{\sigma} &=& (5.386 \pm 0.025)\ \mathrm{ns}\nonumber\\
G_{\sigma} &=& (5.307 \pm 0.032)\ \mathrm{ns}\\
H_{\sigma} &=& 1.586 \pm 0.020.\nonumber
\end{eqnarray}
Fitting the arrival time distribution with a $\Gamma$-pdf partially cuts off the long tail of particles arriving with very high delay. Since the radio emission is, however, dominated by the bulk of the particles, the effect is negligible for our calculations.

\begin{figure}
   \psfrag{R0m}[c][B]{radial distance from shower core [m]}
   \psfrag{t0ns}[c][t]{$t$~[ns]}
   \includegraphics[width=8.6cm]{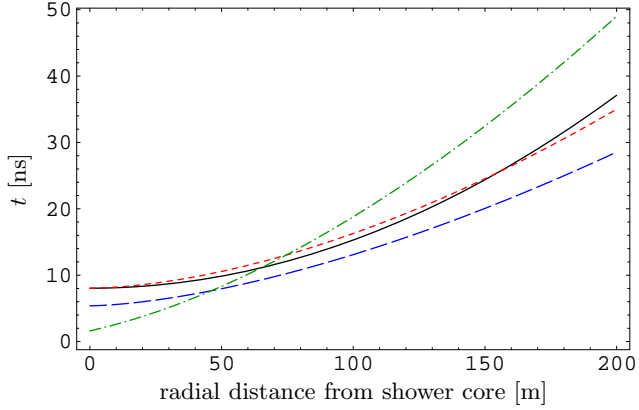}
   \caption{Radial dependence of the particle arrival time distribution. Solid: shower curvature as given by a spherical surface with $K=2,300$~m, short-dashed: $<\!t\!>\!(r)$ as given by \protect\citet{AgnettaAmbrosioAramo1997}, long-dashed: $\sigma_{t}(r)$ as given by \protect\citet{AgnettaAmbrosioAramo1997}, dash-dotted: $\sigma_{t,\mathrm{L}}(r)$ as given by \protect\citet{Linsley1986}.
   \label{fig:curvature_plot}}
   \end{figure}

The thickness of the shower ``pancake'' is directly determined by $\sigma_{t}$. The ``effective curvature'' of the shower front is governed by two factors. On the one hand, there is a delay of the \begin{em}first\end{em} particles of the $\Gamma$-pdf arriving at distance $r$ from the shower core with respect to the first particles arriving in the shower centre. This effect is not included in the \citet{AgnettaAmbrosioAramo1997} data. Here we assume that the delay is negligible for the shower distances $\lesssim 100$~m that we are interested in (the first particles can, with good agreement, be assumed to lie on a flat surface). On the other hand, the \begin{em}mean\end{em} particle delay rises as one goes to greater distances from the shower core, a fact represented by the increase of $<\!\!t\!\!>\!(r)$. The shower curvature determined by $<\!\!t\!\!>\!(r)$ can be expressed very well with a spherical surface of curvature radius $K=2,300$~m, as can be seen in Figure \ref{fig:curvature_plot}.

For comparison, we also examine the \citet{Linsley1986} parametrisation for $\sigma_{t}$, which is defined as
\begin{equation}
\sigma_{t,L}=G_{\mathrm{L}}\ (1+r/r_{\mathrm{L}})^{H_{\mathrm{L}}},
\end{equation}
where, for a $10^{17}$~eV vertical air shower, we have $G_{\mathrm{L}}=1.6$~ns, $r_{\mathrm{L}}=30$~m and $H_{\mathrm{L}}=1.68 \pm 0.14$. As \citet{Linsley1986} does not specify the functional form of the arrival time distribution, we assume that it also corresponds to a $\Gamma$-pdf. However, since $<\!t\!>$ and $\sigma_{t}$ are not independent in this parametrisation, we have to assume a $<\!t\!>(r)$ that fits the $\sigma_{t}(r)$-distribution given by Linsley. From the fact that
\begin{equation}
<\!t\!>=\sqrt{1+B}\ \sigma_{t}
\end{equation}
and the relative constancy of $B(r)$ in the \citet{AgnettaAmbrosioAramo1997} data, a natural choice for the distribution is given by
\begin{equation}
<\!t\!>_{\mathrm{L}}\!(r)=\frac{<\!t\!>_{\mathrm{A}}\!(r)}{\sigma_{t,\mathrm{A}}(r)}\ \sigma_{t,\mathrm{L}}(r),
\end{equation}
where the index A refers to the Agnetta parametrisations and L refers to the Linsley parametrisations.

\subsection{Particle energy distribution}

The average energy of the electrons and positrons in an air shower corresponds to $\sim 30$~MeV, i.e.\ $\gamma \sim 60$ \citep{Allan1971}. As a very crude approximation, one can therefore adopt a mono-energetic configuration of particles with $\gamma \equiv 60$. To illustrate the effects induced by a more realistic particle energy distribution, we compare this with a (spatially uniform) broken power-law distribution of particle energies where $\mathrm{d}N/\mathrm{d}\gamma$ rises linearly with $\gamma$, peaks at $\gamma_{0}=60$ and then declines as $\gamma^{-2}$:
\begin{equation}
p(\gamma)=\left(\frac{\gamma}{\gamma_{1}}\right)^u \left(1-\mathrm{e}^{-(\gamma / \gamma_{1})^{w - u}}\right),
\end{equation}
where we set $u=1$, $w=-2$ and $\gamma_{1}=74.2$ which corresponds to a peak of the distribution at $\gamma_{0}=60$. One can then calculate the emission of an ``energy averaged'' particle through
\begin{equation}\label{eqn:gamaverage}
\vec{E}_{\gamma}(\vec{R},\omega) = p_{0} \int_{\gamma_{\mathrm{min}}}^{\gamma_{\mathrm{max}}}p(\gamma)\ \vec{E}(\vec{R},\omega)\ \mathrm{d}\gamma,
\end{equation}
where the normalisation constant $p_{0}$ is
\begin{equation}
p_{0}=\frac{1}{\int_{\gamma_{\mathrm{min}}}^{\gamma_{\mathrm{max}}}p(\gamma)\ \mathrm{d}\gamma}.
\end{equation}
This energy integration leaves the number of particles unchanged. Note, however, that the total amount of energy in the particles varies with changing $\gamma_{\mathrm{min}}$ and $\gamma_{\mathrm{max}}$. Additionally, $\gamma_{\mathrm{min}}$ must not be chosen too small as our derivations include approximations that are only valid in the ultra-relativistic case. In general, the presence of high-energy particles amplifies the emission near the shower centre, whereas low-energy particles enhance the radiation at high distances due to their wider beaming cone.

We will continue to compare results with both mono-energetic and broken power-law particle distributions and differentiate the two cases through the absence or presence of an additional index $\gamma$. Any result calculated for a broken power-law distribution, indicated through an index $\gamma$, also applies to the mono-energetic case if one substitutes the energy-averaged $\vec{E}_{\gamma}(\vec{R},\omega)$ by the original $\vec{E}(\vec{R},\omega)$.

\section{Coherence: longitudinal effects}

Having established the emission from individual particles and the spatial distribution of particles in the air shower, we can now calculate the emission from the shower maximum, taking into account the inherent (observer-frame) shower structure. The phase differences between the pulses from the individual particles lead to strong coherence effects that significantly change the spectrum of the received emission from that of a fully coherent synchrotron pulse. (For a general discussion of coherence effects regarding synchrotron radiation see also \citealt{AloisioBlasi2002}.)

To get a better understanding of the spectral features, we start with a strongly simplified configuration that neglects any lateral structure: We reduce the shower to its core. In this approximation, the air shower ``pancake'' of thickness $d$ is collapsed to a one-dimensional line of length $d$. The charges distributed along the line are adopted to move synchronously, i.e.\ the momentum distribution of the particles at a given time corresponds to a $\delta$-function. The emission from a particle at distance $\vec{R}=\hat{\vec{n}} R$ from the observer is given by $\vec{E}(\vec{R},\omega)$ as defined in Eq.\ (\ref{eqn:Epairpart}). For a particle offset by a distance $x$ from the line-centre (located at $\vec{R_{0}}$) along the shower core direction $\vec{\hat{l}}$, we therefore have
\begin{eqnarray}
\vec{E}(\vec{R_{0}}+x \hat{\vec{l}},\omega) &\propto& \frac{1}{|\vec{R_{0}}+x \hat{\vec{l}}|}\ \exp\left(\frac{\mathrm{i} \omega |\vec{R_{0}}+x \hat{\vec{l}}|}{c}\right)A_{\parallel}(\omega)\nonumber\\
&\approx& \frac{1}{|\vec{R}_{0}|} \exp\left(\frac{\mathrm{i} \omega |\vec{R_{0}}+x {\hat{\vec{n}}}|}{c}\right)A_{\parallel}(\omega)\nonumber\\
&=& \frac{1}{R_{0}} \exp\left(\frac{\mathrm{i} \omega (R_{0}+x) }{c}\right)A_{\parallel}(\omega).
\end{eqnarray}
Here we keep the distance of the particle at the constant value $R_{0}$ for the first factor, which only introduces negligible errors since $d \ll R_{0}$. The approximation of $\hat{\vec{l}} \approx \hat{\vec{n}}$ in the second factor is justified since $A_{\parallel}$ only gives significant contributions if the directions of $\hat{\vec{l}}$ and $\hat{\vec{n}}$ enclose angles of order $\gamma^{-1}$ or smaller. In other words, projection effects do not play a significant role because only in the regime where they are very small, we have significant emission from the particles.

Defining the particle distribution function $f(x)$ such that $\int_{-\infty}^{+\infty}f(x)\ \mathrm{d}x=1$, and taking into account the particle energy distribution, the integrated emission from $N$ particles is then given by
\begin{eqnarray}
\vec{E}_{\gamma,l}^{N}(\vec{R},\omega) &=& \int_{-\infty}^{+\infty}N f(x)\ \vec{E}_{\gamma}(\vec{R_{0}}+x \vec{\hat{l}},\omega)\  \mathrm{d}x\nonumber\\
&\approx& N \vec{E}_{\gamma}(\vec{R_{0}},\omega) \int_{-\infty}^{+\infty} f(x)\ \mathrm{e}^{\mathrm{i}\omega\frac{x}{c}}\ \mathrm{d}x\nonumber\\
&=& N \vec{E}_{\gamma}(\vec{R}_{0},\omega)\ S(\omega).
\end{eqnarray}
Note that this basically corresponds to a Fourier transformation, i.e.\ the function $S(\omega$) modulating the field strength spectrum is given by the Fourier transform of the particle distribution function, as in standard diffraction theory. (The energy spectrum is then modulated by $|S(\omega)|^{2}$.)

We will now compare a number of different distributions of particles along the line to better understand the coherence effects that arise from longitudinal distributions of particles.

\subsection{Uniform line charge}

The easiest case of a line charge is a uniform distribution of particles along a line of length $d$,
\begin{equation}
f(x)=\left\{\begin{array}{ll}1/d & \quad\textrm{for} \quad|x|\leq d/2\\0 & \quad\textrm{for} \quad|x|>d/2\end{array}\right..
\end{equation}
Integration over $x$ then leads to the well-known $(\sin z/z)^{2}$ modulation of the energy spectrum that corresponds to the diffraction pattern of a rectangular opening, 
\begin{equation}
S(\omega)=\int_{-\frac{d}{2}}^{+\frac{d}{2}}\frac{1}{d}\ \mathrm{e}^{\mathrm{i}\omega\frac{x}{c}}\ \mathrm{d}x=\frac{\sin{(d \omega / 2c})}{d \omega / 2c}.
\end{equation}

\subsection{Gaussian line charge}

A more realistic case is that of a Gaussian distribution of particles along the line. The width of the distribution is set by the standard deviation of the Gaussian $\sigma$ (the FWHM then corresponds to $\sqrt{4\ \ln{4}}\ \sigma \approx 2.35\ \sigma$), with the distribution being defined as
\begin{equation}
f(x)=\frac{1}{\sigma \sqrt{2\pi}}\ \exp\left(-\frac{1}{2}\frac{x^{2}}{\sigma^{2}}\right).
\end{equation}
The coherence function then equals
\begin{equation}
S(\omega)=\exp\left(-\frac{1}{2}\frac{\sigma^{2}}{c^{2}}\omega^{2}\right),
\end{equation}
i.e.\ a Gaussian as well, which is clear from the fact that the Fourier transform of a Gaussian is again a Gaussian.

\subsection{Asymmetrical $\Gamma$-distribution}

A realistic longitudinal particle distribution is given by an arrival-time distribution as specified by Eq.\ (\ref{eqn:gampdf}) with the substitution $x=ct$,
\begin{equation}
f(x)=\left\{\begin{array}{ll} A \left(\frac{x}{c}\right)^{B}\ \exp\left(-C\ \frac{x}{c}\right) & \quad \textrm{for} \quad x>0 \\ 0 & \quad \textrm{for} \quad x \leq 0\end{array}\right.,
\end{equation}
where from the normalisation of $f(x)$ follows
\begin{equation}
A=\left[C^{-(1+B)} \Gamma(1+B)\right]^{-1}.
\end{equation}
The corresponding coherence function $S(\omega)$ is then given by
\begin{eqnarray}
S(\omega)&=& \left(1+\frac{\omega^{2}}{C^{2}}\right)^{-\frac{1}{2}(1+B)} \exp\left[\mathrm{i}(1+B) \arctan\left(\frac{\omega}{C}\right)\right]\nonumber\\
&\times& \exp\left(-\mathrm{i} \omega <\!t\!>_{\mathrm{A}}\right),
\end{eqnarray}
where the last phase factor is needed to ``centre'' the asymmetrical distribution on the curved shower surface to make it comparable to the symmetrical uniform and Gaussian distribution for the later calculations taking into account lateral structure. (The origins of the $\Gamma$-pdfs then again lie on a flat surface as discussed in Section \ref{sec:shower:arrivalt}.)

\subsection{Model calculations}\label{sec:longcoherence}

The results derived so far allow us to perform a number of model calculations in order to analyse the effects of longitudinal particle distributions on the observed spectra as well as the dependence of the emission on the observer's radial distance from the shower core. Where no analytical result was presented, integrations and other calculations are done numerically. We model the maximum of a vertical air shower with primary particle energy $E_{\mathrm{p}} = 10^{17}$~eV and therefore $N=10^{8}$ particles at a height of $R_{0}=4$~km. This is a realistic value as outlined in Section \ref{sec:shower:longdev}. The earth's magnetic field is adopted with a strength of $B=0.3$~G and, for simplicity and symmetry reasons, assumed to be perpendicular to the shower core and thus parallel to the earth's surface (a realistic value for Central Europe would be $B=0.5$~G with declinations around 70$^{\circ}$).

The thickness of the air shower ``pancake'' is determined by the standard deviation $\sigma_{t}$ as parametrised in Sect.\ \ref{sec:shower:arrivalt}. To ensure an equivalent width of the $\Gamma$-pdf, the uniform line charge and the Gaussian line charge, we normalise the latter distributions such that they have a standard deviation of $c\sigma_{t}$. For the Gaussian distribution, this corresponds to $\sigma=c\sigma_{t}$. The uniform line charge must be set to a total length of $d=2 \sqrt{3} c\sigma_{t}$. Evaluated in the shower core, $c\sigma_{t}(0)=1.61$~m, which results in $d=5.6$~m.

   \begin{figure}
   \psfrag{nu0MHz}[c][B]{$\nu$~[MHz]}
   \psfrag{Eomega0muVpmpMHz}[c][t]{$\left|\vec{E}(\vec{R},\omega)\right|$~[$\mu$V~m$^{-1}$~MHz$^{-1}$]}
   \includegraphics[width=8.6cm]{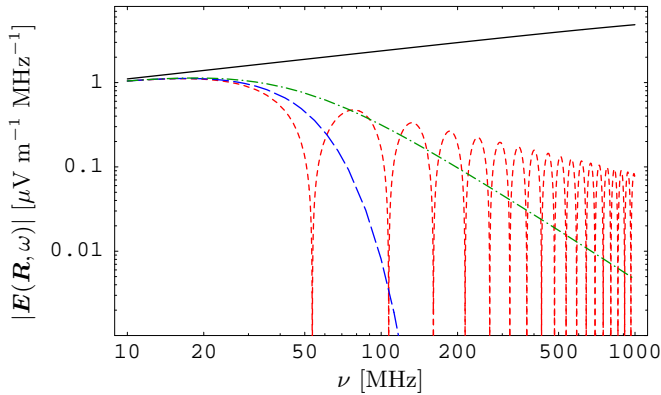}
   \caption{
   \label{fig:coherence_variants_linecharge}
   $\left|\vec{E}(\vec{R},2\pi\nu)\right|$-spectrum in the centre of the area illuminated by the maximum of a $10^{17}$~eV air shower with $R_{0}=4$~km and $\gamma\equiv60$. Solid: full coherence, short-dashed: uniform 5.6~m line charge, long-dashed: Gaussian line charge with $\sigma=1.61$~m, dash-dotted: asymmetrical $\Gamma$-distribution with c$\sigma_{t}=1.61$~m
   }
   \end{figure}

Fig.\ \ref{fig:coherence_variants_linecharge} compares the spectral modulations arising from the different longitudinal particle distributions. If the particles radiated fully coherently --- i.e.\ moved ``as one particle'' on the exact same trajectory --- the field strength spectrum of the emission would simply be a synchrotron spectrum enhanced by a factor $N$. The coherence effects modulate this spectrum by the coherence function $S(\omega)$. In the case of the uniform line charge, we see the first coherence minimum at $\approx 54$~MHz, which corresponds to $c/d$. The Gaussian line charge spectrum does not exhibit such a sharp minimum, but is strongly attenuated at higher frequencies. The asymmetrical $\Gamma$-pdf lies between these two simplified models.

Obviously the longitudinal effects very strongly modulate the emitted spectrum at high frequencies ($\gtrsim 50$~MHz) and therefore are an important limiting factor for the choice of a suitable observing frequency. The thickness of the air shower ``pancake'' has a very profound and direct influence on the emitted radiation and could therefore be probed directly through observations of radio emission from EAS at frequencies $>50$~MHz.

   \begin{figure}
   \psfrag{R0m}[c][B]{distance from shower centre [m]}
   \psfrag{Eomega0muVpmpMHz}[c][t]{$\left|\vec{E}(\vec{R},\omega)\right|$~[$\mu$V~m$^{-1}$~MHz$^{-1}$]}
   \includegraphics[width=8.6cm]{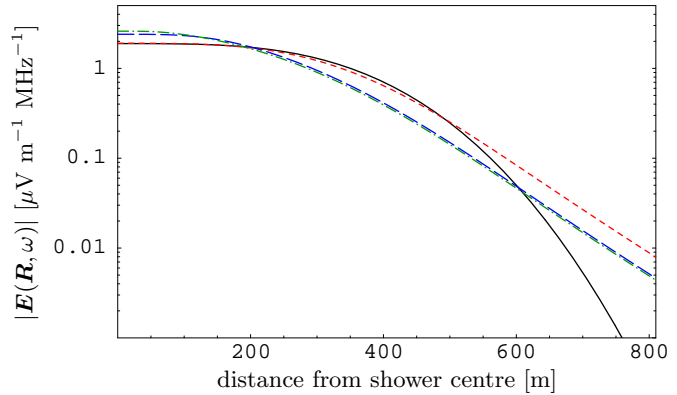}
   \caption{Radial dependence of $\left|\vec{E}(\vec{R},2\pi\ 50\ \mathrm{MHz})\right|$ for the maximum of a $10^{17}$~eV point source shower with $R_{0}=4$~km for the $\gamma\equiv60$ case (solid) and for broken power-law distributions from $\gamma=5$--120 (short-dashed), $\gamma=5$--1000 (long-dashed) and $\gamma=5$--10000 (dash-dotted).
   \label{fig:radius_dependence_linecharge}}
   \end{figure}

Another interesting characteristic of the radiation is its radial dependence at a given frequency as illustrated by Fig.\ \ref{fig:radius_dependence_linecharge} for the case of full coherence without any longitudinal distribution, i.e.\ for particles concentrated in a point source. In this case, the associated emission pattern is purely governed by the inherent emission pattern of the synchrotron pulses. The extent of the illuminated area on the ground is an important characteristic that ultimately limits the probability to detect very scarce ultra-high energy cosmic ray air showers with a given collecting area.

As expected, adoption of the broken power-law distribution of particle energies influences the radial emission pattern. Going from the mono-energetic case to the broken power-law energy distribution mainly amplifies the emission in the centre region due to the presence of high energy particles that radiate more strongly, but into a smaller beaming cone. At the same time, the low-energy particles amplify the emission at very high distances due to their wider emission pattern. The drop in the number of medium-energy particles is correspondingly reflected in a drop of the emission on medium scales. Obviously, there is only negligible difference when increasing the upper limit $\gamma_{\mathrm{max}}$ from a value of 1000 to higher values such as 10000. For the remaining calculations, we therefore adopt a distribution with $\gamma$ in the range 5--1000 to minimise computation time.

\section{Coherence: lateral effects}\label{sec:lateffects}

After having analysed coherence effects arising from longitudinal distributions of particles, we now take a look at the influence of the lateral structure of the air shower on the radio emission. This we accomplish by ``smearing out'' the line charge considered so far over a segment of a spherical surface with (for the moment) constant thickness $d$. Inside this ``shell'' we continue to consider the types of longitudinal particle distributions introduced in Section \ref{sec:longcoherence}.

\subsection{Geometry}

   \begin{figure}
   \includegraphics[width=8.6cm]{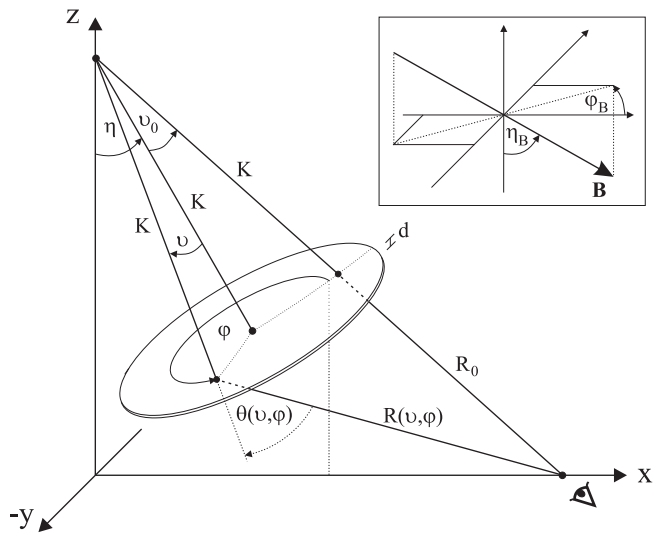}
   \caption{
   \label{fig:geometrie_skizze}
   The geometry for the air shower maximum.
   }
   \end{figure}

The geometry of the air shower maximum is defined as in Fig.\ \ref{fig:geometrie_skizze} and characterised by the curvature radius of the shower surface $K$, the shower shell thickness $d$ and the shower inclination angle $\eta$. The observer is positioned on the $x$-axis at a minimum distance $R_{0}$ from the shower surface, with an inclination angle $\vartheta_{0}$ to the shower core. The magnetic field strength $B$, inclination $\eta_{B}$ and azimuthal direction $\varphi_{B}$ determine the configuration of the earth's magnetic field.

To derive the total emission from the air shower maximum, we now have to integrate over the shell and hence must relate $\vec{E}(\vec{R},\omega)$ and consequently the quantities going into $\vec{E}(\vec{R},\omega)$ to the position on the surface as given by $\vartheta$ and $\varphi$. We refer the reader to the appendix for the details of these calculations.

\subsection{Approximations} \label{sec:reducedtheta}

In order to facilitate the integration over the shower shell we apply a number of approximations. First, we sum the contributions from the different regions of the air shower maximum in a scalar way, i.e.\ we do not evaluate Eq.\ (\ref{eqn:epardir}) but set $\vec{\hat{\mathrm{e}}}_{\parallel}(\vartheta,\varphi) \equiv \vec{\hat{\mathrm{e}}}_{\parallel}(0,0)$, which is justified due to the minute changes in the direction of $\vec{\hat{\mathrm{e}}}_{\parallel}(\vartheta,\varphi)$ over the shower surface. The general polarisation direction of the radiation then is perpendicular to both the shower axis and the magnetic field direction. Second, as pointed out in the appendix, we assume that the instantaneous velocity vectors of the particles in the shower shell are perpendicular to the shower surface at the moment corresponding to the origin of Figure \ref{fig:jackson_skizze}. This again corresponds to a $\delta$-function for the distribution of the particle momenta, and in this strict sense, the minimum angle to the line-of-sight $\theta(\vartheta,\varphi)$ is given by Eq.\ (\ref{eqn:costheta}).

Adoption of this $\theta$, however, yields a very conservative estimate for the emission. $\theta(\vartheta,\varphi)$ as calculated by Eq.\ (\ref{eqn:costheta}) overestimates the minimum angle to the line-of-sight as defined in Fig.\ \ref{fig:jackson_skizze} in case of a particle trajectory bending towards or away from the observer, where a significantly reduced $\theta$ is reached during a later or earlier position on the particle trajectory. The amount of ``compensation'' in $\theta$ attainable by this effect is considerable since the ratio of mean free path length to curvature radius of $\gamma=60$ electrons is $\sim 450\,\mathrm{m}/3400\,\mathrm{m}\approx 8\,\gamma^{-1}$. $\theta$ is therefore effectively reduced to its irreducible component given by
\begin{equation}\label{eqn:costhetareduced}
\sin \theta(\vartheta,\varphi) = \vec{\hat{B}} \cdot \vec{\hat{n}}(\vartheta,\varphi).
\end{equation}
Adoption of this value for $\theta(\vartheta,\varphi)$ yields a more realistic estimate of the emission from the air shower shell, and at the same time takes into account the asymmetry of the emission pattern in $\varphi$ that arises from the magnetic field configuration. Without a more precise criterion for the maximum compensation achievable, however, the radial dependence of the emission pattern at very high distances is obviously no longer valid. We therefore continue to work with both the conservative approach using Eq.\ (\ref{eqn:costheta}) and the ``reduced $\theta$'' definition in order to compare the two cases.

The change in $R$ associated to the adoption of the ``optimum position'' on the particle trajectory is negligible because of the following reasons:
\begin{itemize}
\item{the compensated angles are small, therefore the changes in $R$ are small}
\item{additional attenuation/amplification through the $1/R$-term is thus negligible} 
\item{there is no significant change of phase since the particle velocity $v \approx c$ and the trajectory is only mildly curved.}
\end{itemize}

\subsection{Integration} \label{sec:integration}

Using these geometrical relations, the integrated spectrum of the emission from the air shower maximum with $N$ particles can be calculated as
\begin{eqnarray}
\vec{E}_{\gamma,\mathrm{S}}^{N}(\omega) &=& \rho_{0}\ \int_{0}^{2\pi}\!\mathrm{d}\varphi \int_{0}^{{r_{\mathrm{M}}/K}}\!\mathrm{d}\vartheta\ K^{2} \sin\vartheta \nonumber\\
&\times& \rho_{\mathrm{NKG}}(r(\vartheta,\varphi))\ \vec{E}_{\gamma}(\vec{R}(\vartheta,\varphi),\omega)
\end{eqnarray}
with the normalisation factor
\begin{equation}
\rho_{0}=N\left[\int_{0}^{2\pi}\!\!\!\!\mathrm{d}\varphi \int_{0}^{{r_{\mathrm{M}}/K}}\!\!\!\!\mathrm{d}\vartheta\ K^{2} \sin\vartheta\  \rho_{\mathrm{NKG}}(r(\vartheta,\varphi))\right]^{-1}\!\!\!\!.
\end{equation}
Cutting off the integration at $\vartheta=r_{\mathrm{M}}/K$ significantly reduces computation time while giving acceptable accuracy as $\gtrsim 80$~\% of the particles are included in this region. The remaining particles are redistributed over the integration region by the normalisation according to the NKG-profile, which might lead to a slight overestimation of the emission strength near the shower centre.

\subsection{Model Calculations}

We again examine the frequency and radial dependence of the emission to study the effects introduced through the lateral particle distribution. The basic parameters adopted are the same as in Sec.\ \ref{sec:longcoherence}, but we assume a broken power-law particle energy distribution from $\gamma=5$--1000 for all calculations. The curvature radius of the shell is adopted as $K=2,300$~m and the Moli\`ere radius set to $r_{\mathrm{M}}=117$~m as discussed in Sects.\ \ref{sec:shower:arrivalt} and \ref{sec:lateraldistri}, correspondingly.

\begin{figure}
   \psfrag{nu0MHz}[c][B]{$\nu$~[MHz]}
   \psfrag{Eomega0muVpmpMHz}[c][t]{$\left|\vec{E}(\vec{R},\omega)\right|$~[$\mu$V~m$^{-1}$~MHz$^{-1}$]}
   \includegraphics[width=8.6cm]{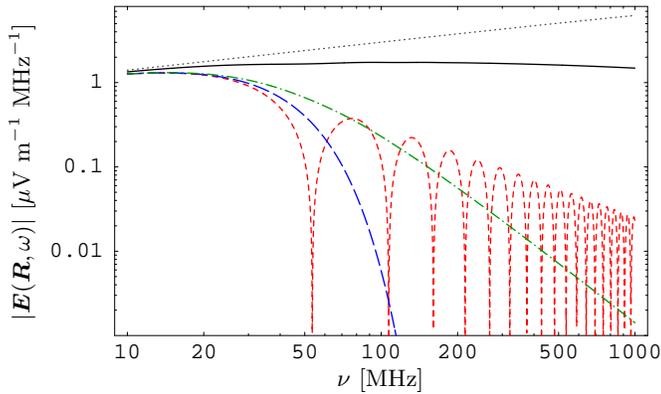}
   \caption{
   \label{fig:coherence_variants_realistic}
   $\left|\vec{E}(\vec{R},2\pi\nu)\right|$-spectrum at the centre of the area illuminated by the maximum of a $10^{17}$~eV air shower with realistic lateral distribution, $R_{0}=4$~km and a broken power-law energy distribution from $\gamma=5$--1000. Solid: full longitudinal coherence, short-dashed: uniform 5.6~m longitudinal distribution, long-dashed: Gaussian longitudinal distribution with $\sigma=1.61$~m, dash-dotted: longitudinal $\Gamma$-distribution with $c\sigma_{t}=1.61$~m. For comparison: fully coherent case without lateral distribution (dotted)
   }
   \end{figure}

In Fig.\ \ref{fig:coherence_variants_realistic} we plot the spectrum received by an observer in the centre of the area illuminated by the air shower maximum, considering the same set of longitudinal particle distributions as before. The spectra look very similar to those of a line charge, but are attenuated additionally at high frequencies.

\begin{figure}
   \psfrag{R0m}[c][B]{distance from shower centre [m]}
   \psfrag{Eomega0muVpmpMHz}[c][t]{$\left|\vec{E}(\vec{R},\omega)\right|$~[$\mu$V~m$^{-1}$~MHz$^{-1}$]}
   \includegraphics[width=8.6cm]{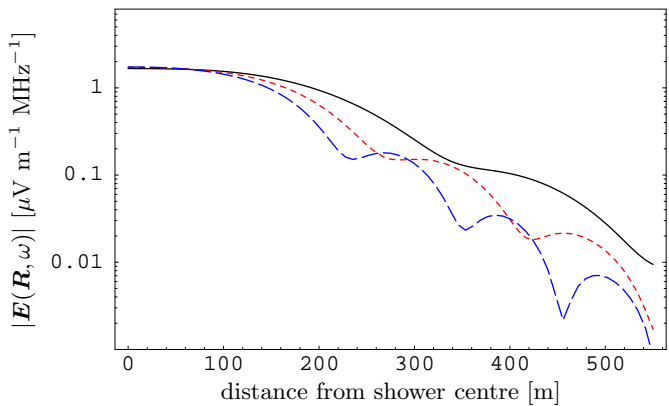}
   \caption{Radial dependence of $\left|\vec{E}(\vec{R},2\pi\nu)\right|$ for the maximum of a $10^{17}$~eV air shower with full longitudinal coherence, realistic lateral structure, ``conservative $\theta$'' approach, $R_{0}=4$~km and a broken power-law energy distribution from $\gamma=5$--1000. Solid: $\nu=50$~Mhz, short-dashed: $\nu=75$~Mhz, long-dashed: $\nu=100$~Mhz
   \label{fig:radius_dependence_realistic}}
   \end{figure}

A more interesting result is illustrated by Fig.\ \ref{fig:radius_dependence_realistic} which demonstrates the effect of a purely lateral particle distribution on the radial dependence of the emission in the ``conservative $\theta$'' scenario, completely ignoring any longitudinal effects. The lateral structure introduces a modulation of the radial dependence, caused by interference of emission from opposite ends of the shower ``disk''. For higher frequencies and, correspondingly, shorter wavelengths, the interference minima move to smaller radial distances.

\begin{figure}
   \psfrag{R0m}[c][B]{distance from shower centre [m]}
   \psfrag{Eomega0muVpmpMHz}[c][t]{$\left|\vec{E}(\vec{R},\omega)\right|$~[$\mu$V~m$^{-1}$~MHz$^{-1}$]}
   \includegraphics[width=8.6cm]{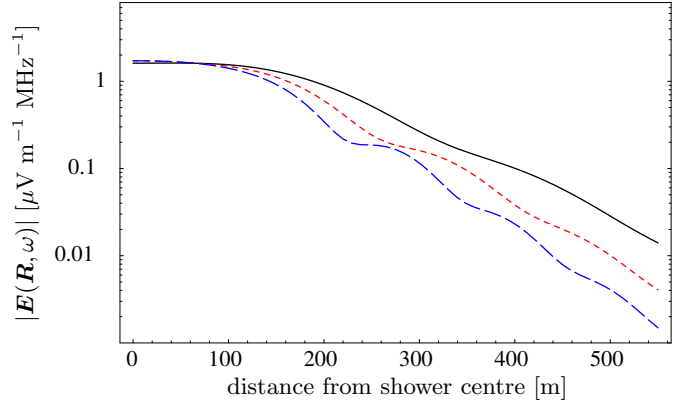}
   \caption{Same as Fig. \ref{fig:radius_dependence_realistic} but with ``reduced $\theta$'' and magnetic field parallel to the direction of the observer.
   \label{fig:raddep_reduced0}}
   \end{figure}

\begin{figure}
   \psfrag{R0m}[c][B]{distance from shower centre [m]}
   \psfrag{Eomega0muVpmpMHz}[c][t]{$\left|\vec{E}(\vec{R},\omega)\right|$~[$\mu$V~m$^{-1}$~MHz$^{-1}$]}
   \includegraphics[width=8.6cm]{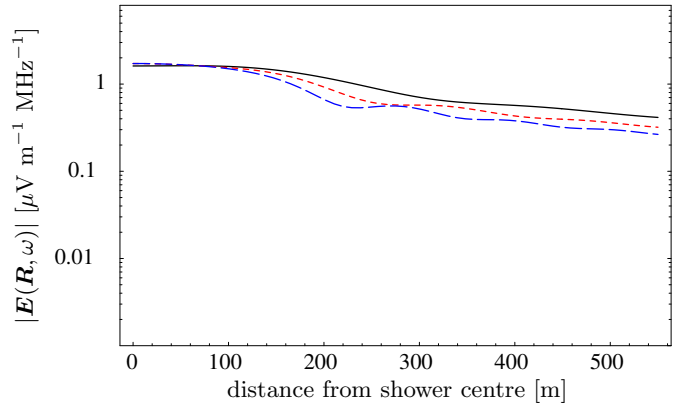}
   \caption{Same as Fig. \ref{fig:radius_dependence_realistic} but with ``reduced $\theta$'' and magnetic field perpendicular to the direction of the observer.}
   \label{fig:raddep_reduced90}
   \end{figure}

In comparison, Figs.\ \ref{fig:raddep_reduced0} and \ref{fig:raddep_reduced90} show the radial dependence in case of the ``reduced $\theta$'' calculations. The interference effects are somewhat washed out and the overall emission level is higher. As expected, there is a drastic asymmetry between the emission pattern in the directions parallel and perpendicular to the magnetic field. In case of Fig. \ref{fig:raddep_reduced90}, where the observer is positioned in a direction perpendicular to the magnetic field, $\theta$ is basically reduced to zero even for distances of a few hundred metres ($\theta \lesssim 8\,\gamma^{-1}$ as explained in Section \ref{sec:reducedtheta}). Correspondingly, the emission pattern is only very slightly attenuated even at high distances.

\section{Flaring disk}

We now combine the results derived so far to obtain a more realistic model of the emission from the maximum of an extensive air shower: a ``flaring'' disk. In other words, we adopt the same geometry as specified in Sec.\ \ref{sec:lateffects}, but now vary the thickness of the disk as a function of position $(\vartheta,\varphi)$ on the shower surface in the form of the varying asymmetric $\Gamma$-pdfs parametrised as in Section \ref{sec:shower:arrivalt}. This geometry therefore correctly takes into account the curvature, the lateral and the longitudinal structure of the air shower maximum.

\begin{figure}
   \psfrag{nu0MHz}[c][B]{$\nu$~[MHz]}
   \psfrag{Eomega0muVpmpMHz}[c][t]{$\left|\vec{E}(\vec{R},\omega)\right|$~[$\mu$V~m$^{-1}$~MHz$^{-1}$]}
   \includegraphics[width=8.6cm]{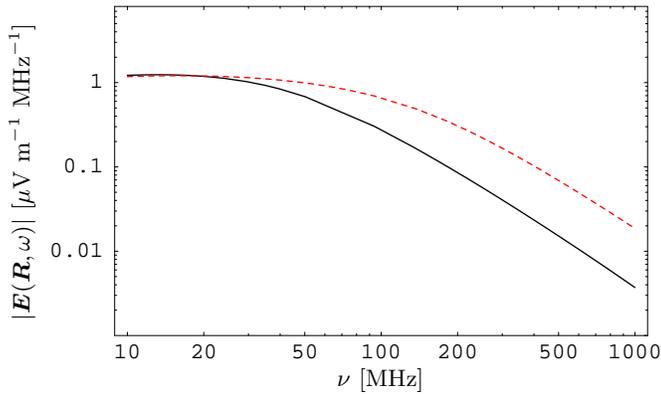}
   \caption{
   \label{fig:coherence_variants_flaring}
   $\left|\vec{E}(\vec{R},2\pi\nu)\right|$-spectrum at the centre of the area illuminated by the maximum of a $10^{17}$~eV air shower with flaring $\Gamma$-pdf, $R_{0}=4$~km and a broken power-law energy distribution from $\gamma=5$--1000. Solid: flaring \protect\citet{AgnettaAmbrosioAramo1997} lateral distribution, short-dashed: flaring \protect\citet{Linsley1986} lateral distribution
   }
   \end{figure}

Fig.\ \ref{fig:coherence_variants_flaring} again shows the spectrum emitted by the air shower maximum as a realistically flaring disk according to the \citet{AgnettaAmbrosioAramo1997} and \citet{Linsley1986} parametrisations. As expected, the spectrum emitted by the Linsley flaring disk extends to higher frequencies than the one generated by the Agnetta flaring disk because of the lower thickness in the shower centre where most of the particles reside (cf.\ Fig.\ \ref{fig:curvature_plot}).

\begin{figure}
   \psfrag{R0m}[c][B]{distance from shower centre [m]}
   \psfrag{Eomega0muVpmpMHz}[c][t]{$\left|\vec{E}(\vec{R},\omega)\right|$~[$\mu$V~m$^{-1}$~MHz$^{-1}$]}
   \includegraphics[width=8.6cm]{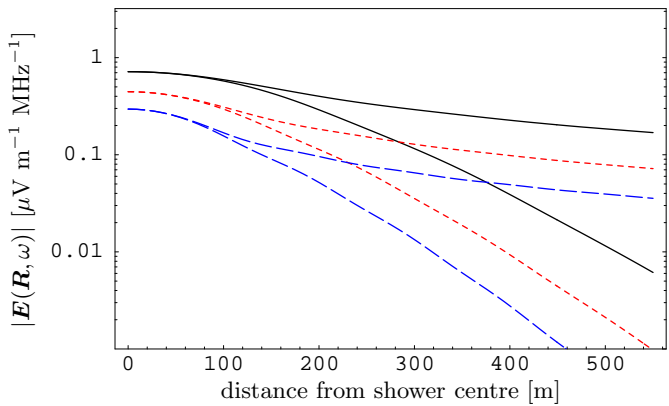}
   \caption{Radial dependence of $\left|\vec{E}(\vec{R},2\pi\nu)\right|$ for the maximum of a $10^{17}$~eV air shower with flaring \protect\citet{AgnettaAmbrosioAramo1997} $\Gamma$-pdf, $R_{0}=4$~km and a broken power-law energy distribution from $\gamma=5$--1000. Solid: $\nu=50$~Mhz, short-dashed: $\nu=75$~Mhz, long-dashed: $\nu=100$~Mhz, upper/lower curves for ``reduced $\theta$'' perpendicular/parallel to magnetic field direction
   \label{fig:radius_dependence_flaring}}
   \end{figure}

The radial dependence at different frequencies is once again shown in Figure \ref{fig:radius_dependence_flaring}. Comparison with the corresponding diagrams for the purely lateral distribution shown in Figs.\ \ref{fig:radius_dependence_realistic}--\ref{fig:raddep_reduced90} shows that the overall emission level drops as the observing frequency is increased due to the dampening of higher frequencies by the longitudinal particle distribution. Additionally, one can again observe a ``smearing out'' of the interference minima. As a consequence, the ``conservative $\theta$'' and the ``reduced $\theta$'' with observer parallel to the magnetic field calculations yield almost identical results.

   \begin{figure}
   \psfrag{time0ns}[c][B]{t~[ns]}
   \psfrag{Efield0muV0m}[c][t]{$\left|\vec{E}(t)\right|$~[$\mu$V m$^{-1}$]}
   \includegraphics[width=8.6cm]{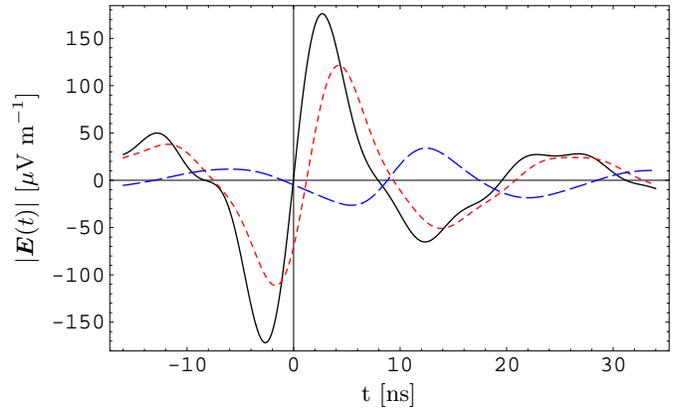}
   \caption{Reconstructed pulses emitted by the maximum of a $10^{17}$~eV shower with flaring \protect\citet{AgnettaAmbrosioAramo1997} $\Gamma$-pdf, broken power-law energy distribution from $\gamma=5$--1000 and $R_{0}=4$~km, using an idealised rectangle filter spanning 40--160~MHz and ``conservative $\theta$'' scenario. Solid: centre of illuminated area, short-dashed: 100~m from centre, dash-dotted: 250~m from centre
   \label{fig:pulsesflaring}}
   \end{figure}

In Fig.\ \ref{fig:pulsesflaring} we have reconstructed the pulse generated by the flaring Agnetta disk as it would be measured by a receiver with a given bandwidth using Eq.\ (\ref{eqn:pulserec}). The pulse amplitude drops noticeably when the observer moves from the centre of the illuminated area on the ground to a distance of 100~m, and is already quite diminished at a distance of 250~m, as expected for the ``conservative $\theta$'' approach. The pulse length of $\approx 8$~ns is a result of the filter bandwidth of 120~MHz, i.e.\ the pulse is bandwidth-limited.

\section{Integration over shower evolution}

The last step in modeling the total air shower emission is to integrate over the shower evolution as a whole. This can be done in a very simplified fashion by approximating the shower evolution with a number of discrete steps. The characteristic scale for these steps is given by the ``radiation length'' of the electromagnetic cascades in air, $X_{0}=36.7$~g~cm$^{-2}$, corresponding to $\approx 450$~m at a height of 4~km. One can therefore discretise the shower evolution into ``slices'' of thickness $X_{0}$, assuming these contain independent generations of particles and therefore radiate independently.

The emission from each of these slices is calculated as that from a flaring disk, taking into account changes of $s$, $R_{0}$, $\vartheta_{0}$, $r_{M}$ and $N$ correctly through the relations given in Section \ref{sec:shower} and reverting to the ``conservative $\theta$'' definition to be able to correctly calculate the emission at great angles. Superposition of the individual slice emissions, correctly taking into account the phases arising from arrival time differences, then leads to the total emission of the shower.

Slices far away from the observer are attenuated both due to the high distance and the decreasing number of particles $N$. The concrete number of far-away slices taken into account is therefore uncritical. The situation is different for the slices close to the observer. In their case, the attenuation through the decreasing number of particles $N$ is more than compensated by the decreasing distance to the observer. In fact, the slices closest to the observer yield the highest contributions of radiation, and the total result depends considerably on the number of nearby slices taken into account. However, at the same time, the illuminated area on the ground, governed by the intrinsic beaming cone, becomes very small for the slices very close to the observer, especially for the high frequencies where the radiation mainly originates from high-energy particles with even smaller beaming cones. Except for low frequency emission in the centre region of the illuminated area, the result for the total emission can therefore be considered robust.

For our vertical $10^{17}$~eV air shower at a height of $R_{0}=4$~km we add the emission from eight slices above and eight slices below the shower maximum to the emission from the maximum itself. The closest slice then lies at $R_{0}=950$~m from the observer, a distance we do not want to fall below because of approximations contained in our calculations that are only valid in the far-field. 

   \begin{figure}
   \psfrag{nu0MHz}[c][B]{$\nu$~[MHz]}
   \psfrag{Eomega0muVpmpMHz}[c][t]{$\left|\vec{E}(\vec{R},\omega)\right|$~[$\mu$V~m$^{-1}$~MHz$^{-1}$]}
   \psfrag{Enu0muVpmpMHz}[c][b]{$\epsilon_{\nu}$~[$\mu$V~m$^{-1}$~MHz$^{-1}$]}
   \includegraphics[width=8.6cm]{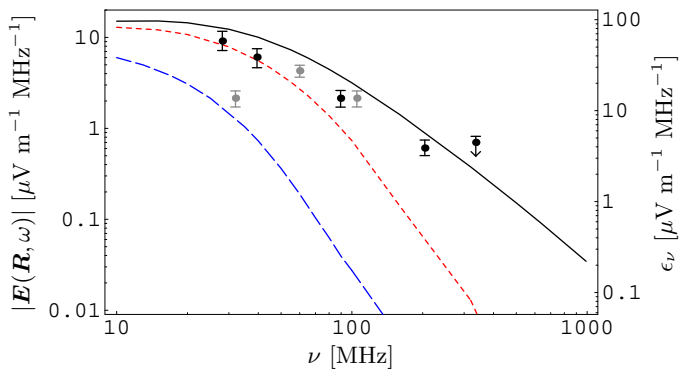}
   \caption{
   \label{fig:spectra_data}
   $\left|\vec{E}(\vec{R},2\pi\nu)\right|$-spectrum of a full $10^{17}$~eV air shower with flaring \protect\citet{AgnettaAmbrosioAramo1997} $\Gamma$-pdf, ``conservative $\theta$'' approach, $R_{0}=4$~km and a broken power-law energy distribution from $\gamma=5$--1000. Solid: centre of illuminated area, short-dashed: 100~m from centre, long-dashed: 250~m from centre, black points: {\em{rescaled}} \protect\citet{Spencer1969} data as presented by \protect\citet{Allan1971}, grey points: {\em{rescaled}} \protect\citet{Prah1971} data
   }
   \end{figure}

   \begin{figure}
   \psfrag{R0m}[c][B]{distance from shower centre [m]}
   \psfrag{Eomega0muVpmpMHz}[c][t]{$\left|\vec{E}(\vec{R},\omega)\right|$~[$\mu$V~m$^{-1}$~MHz$^{-1}$]}
   \psfrag{Enu0muVpmpMHz}[c][b]{$\epsilon_{\nu}$~[$\mu$V~m$^{-1}$~MHz$^{-1}$]}
   \includegraphics[width=8.6cm]{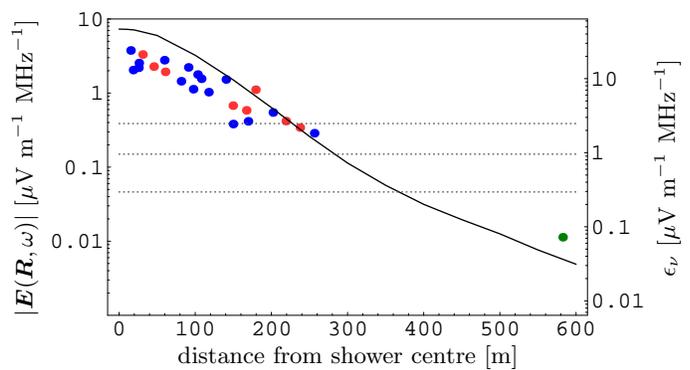}
   \caption{Radial dependence of $\left|\vec{E}(\vec{R},2\pi\ 55\ \mathrm{MHz})\right|$ for a full $10^{17}$~eV air shower with flaring \protect\citet{AgnettaAmbrosioAramo1997} $\Gamma$-pdf, ``conservative $\theta$'' approach, $R_{0}=4$~km and a broken power-law energy distribution from $\gamma=5$--1000, data from \protect\citet{AllanClayJones1970}, horizontal lines from top to bottom: emission strength needed for a $3\sigma$-detection with an individual LOPES antenna or an array of 10 or 100 LOPES antennas
   \label{fig:radius_dependence_flaring_evolution_integrated_data}}
   \end{figure}

   \begin{figure}
   \psfrag{time0ns}[c][B]{t~[ns]}
   \psfrag{Efield0muV0m}[c][t]{$\left|\vec{E}(t)\right|$~[$\mu$V m$^{-1}$]}
   \includegraphics[width=8.6cm]{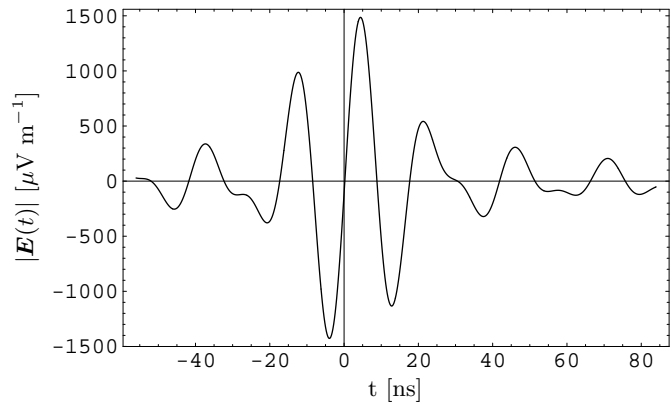}
   \caption{Reconstructed pulse in the centre of the area illuminated by a full $10^{17}$~eV shower with flaring \protect\citet{AgnettaAmbrosioAramo1997} $\Gamma$-pdf, broken power-law energy distribution from $\gamma=5$--1000 and $R_{0}=4$~km, using an idealised rectangle filter spanning 42.5--77.5~MHz
   \label{fig:pulsesflaringfullevolution}}
   \end{figure}

The main effect of the integration over the shower evolution is a boosting of the total emission because of the increased total number of particles taken into account, as can be seen in the spectra shown in Fig.\ \ref{fig:spectra_data}. For frequencies of $\sim 40$~MHz and radial distances of  $\sim 100$~m, the amplification factor corresponds to $\sim 10$. Apart from the overall amplification, the radial dependence is significantly steepened because the important nearby slices only contribute at low radial distances as discussed earlier. This can be seen when comparing Fig.\ \ref{fig:radius_dependence_flaring_evolution_integrated_data} with the earlier results for the ``conservative $\theta$'' case. For illustration purposes, we also present a reconstructed pulse from the shower as a whole as it would be measured by an observer in the centre of the illuminated area using the LOPES bandwidth of 35~MHz in Figure \ref{fig:pulsesflaringfullevolution}.

\section{Discussion}

The calculations presented here represent only a few illustrative examples of possible configurations of EAS that could be calculated with our model. These examples, however, already demonstrate the most important dependencies between shower structure and emission spectrum as well as radial emission pattern.

\subsection{Theoretical results}

As expected, the thickness of the air shower pancake, and correspondingly in our model the width of the longitudinal particle arrival time distributions, is the main factor determining the position of the high-frequency cut-off in the spectrum. Typical longitudinal scales of a few metres lead to frequency cut-offs in the 100~MHz regime, which supports a choice of observing frequency well below 100~MHz. Due to the strong dependence of the spectral cut-off on the shower thickness, radio emission from EAS could be used very effectively to probe the longitudinal structure of air showers during their evolution, a quantity that is not well known at the moment. 

The radial emission pattern is mainly governed by the inherent emission pattern of the synchrotron pulses and the superposition of the beamed emission from different parts of the air shower evolution as a whole. Additionally, the lateral extent of the air shower slightly influences the size of the illuminated area on the ground through the resulting coherence minima. A profound change in the radial emission pattern is visible when one adopts the ``reduced $\theta$'' approach, which predicts significant radio emission up to higher distances depending on the relative orientation of observer and magnetic field. This is an important prerequisite for the detection of ultra-high energy EAS with an array of affordable collecting area in combination with particle detector arrays such as KASCADE Grande or the Pierre Auger Observatory and will be verifiable by LOPES.

The emitted total power in the coherent regime at low frequencies scales as the number of particles squared, which could therefore be probed directly by radio measurements of EAS, yielding information about the primary particle energy.

We have not explicitly presented how variations of other parameters influence the radio emission, but most of the associated effects are fairly straight-forward to foresee: The emitted power scales linearly with $B$-field strength. The declination of the $B$-field in Central Europe effectively decreases the value of $B$ and introduces an asymmetric pattern to the radial dependence. An increase of the primary particle energy will boost the radiation because higher-energy showers will have their maximum closer to the observer. At the same time, the number of particles increases linearly with primary particle energy and the power emitted at low frequencies increases as number of particles squared, which more than compensates the shrinking of the illuminated area on the ground. Inclined air showers will cause an asymmetric emission pattern and an attenuation of the emitted power because they reach their development at higher altitudes. A stronger curvature of the shower front will shift the interference minima to smaller radial distances and thus slightly decrease the effective size of the illuminated area on the ground.

\subsection{Comparison with experimental data}

A number of experiments have clearly established the presence of radio emission from cosmic ray air showers in the past. A dependence of the polarisation of the emitted radiation on the earth's magnetic field direction was also confirmed by a number of experiments (e.g., \citealt{AllanClayJones1969}), supporting the case for the geomagnetic emission mechanism. The actual strength of the emission, however, is still largely unknown at present state. The analysis of \citet{Allan1971} led to a widely used formula summarising the presumed dependencies:
\begin{eqnarray}
\epsilon_{\nu}&=&20\ \mu\mathrm{V}\,\mathrm{m}^{-1}\,\mathrm{MHz}^{-1} \left(\frac{E_{\mathrm{p}}}{10^{17}\ \mathrm{eV}}\right)\nonumber\\
&\times& \sin{\alpha}\ \cos{\eta}\ \exp\left(-\frac{r}{r_{0}(\nu,\eta)}\right),
\end{eqnarray}
where the scale factor $r_{0}$ corresponds to (110$\pm$10)~m at $\nu=55$~MHz and for $\eta<35^{\circ}$. Later works (e.g., \citealt{Sun1975}; \citealt{Prah1971} and references therein), however, yielded values as low as 1--5~$\mu$V~m$^{-1}$~MHz$^{-1}$. A recent experiment in conjunction with the CASA/MIA array conducted by \citet{GreenRosnerSuprun2003} was only able to place upper limits of $\epsilon_{\nu}=31$---34~$\mu$V~m$^{-1}$~MHz$^{-1}$ on the emission strength.

Part of these discrepancies could be explained by uncertainties in the primary particle energy calibration at the time the experiments were made. A number of authors involved in the past works suspect the calibration of the radio measurements to be the major source of uncertainty \citep{AtrashkevichVedeneevAllan1978}. Additionally, the documentation of the available data is not always totally precise regarding the included energy ranges of primary particles, the selection of allowed zenith angles, the radial distance to the shower axis or the back-projection of the electric field vector in the plane normal to the shower axis and earth's magnetic field, which further complicates the issue.

Extremely low values of $\epsilon_{\nu}$ of only $1\ \mu$V~m$^{-1}$~MHz$^{-1}$ or even lower are, however, disfavoured by the fact alone that air showers actually {\em{have been measured}} by experiments with only a few antennas (e.g., two in case of \citealt{Prah1971}) with receivers of only a few MHz bandwidth in the early experiments.

In this difficult situation, we choose to revert to the well documented data of \citet{AllanClayJones1970} as the basis of our analysis. A comparison of these data with our predicted radial dependence of the emission is shown in Figure \ref{fig:radius_dependence_flaring_evolution_integrated_data}. While we clearly overpredict the emission strength in the centre, the general radial dependence fits relatively well. Regarding the spectral dependence, we make use of the \citet{Spencer1969} data as presented, converted and complemented in \citet{Allan1971} as well as the \citet{Prah1971} data. These data sets, again, yield considerably lower values of $\epsilon_{\nu}$, and we manually scale them up to make them consistent with the \citet{AllanClayJones1970} radial data. While the absolute values presented in Fig.\ \ref{fig:spectra_data} therefore are somewhat arbitrary, the trend in the dependence actually does correspond to the spectral dependence that we predict near the shower core.

All in all, we overpredict even the most optimistic past data by a factor $\sim 2$, which is, however, not too surprising considering the very simplified integration over the shower evolution as a whole and the problems involved especially in the centre region. Additionally, the cutoff of the spatial integration as stated in Sec.\ \ref{sec:integration} redistributes further emission to the centre region.

We feel that having achieved a result which is consistent with past experimental data within a factor of ``a few'' using such approximate descriptions of the shower characteristics and a mainly analytical approach incorporating major approximations is a very encouraging outcome. In addition, our result further supports the geomagnetic emission mechanism as the dominant source of radio emission from EAS. A huge improvement of our model will only be possible if we revert to elaborate computer simulations that use fewer approximations, more realistic particle distributions (e.g., by interfacing our model to CORSIKA), and include additional effects such as the charge excess mechanism. Consequently, this will be the next step in our modeling efforts.

Apart from further development towards a more sophisticated model of radio emission from EAS, the most important aim for the near future therefore clearly is the obtainment of new, {\em{reliable}} data --- as will be provided by LOPES, which should be able to easily measure the radio emission from a 10$^{17}$~eV air shower as illustrated by the signal-to-noise levels overplotted in Figure \ref{fig:radius_dependence_flaring_evolution_integrated_data}.

\section{Conclusions}

We have analysed properties of radio emission from EAS in the scenario of coherent geosynchrotron emission. Our step-by-step analysis has helped to disentangle the coherence effects arising from the different physical scales present in the air shower and to get a good feeling for the relative importance of these effects. While the spectral cutoff is directly governed by the longitudinal extent of the air shower, the radial dependence arises from the intrinsic beaming of the synchrotron radiation and its superposition over the shower evolution as a whole.

The emitted radio power is of the expected order of magnitude, which is the strongest constraint we can make at the moment due to the large uncertainties associated with the available experimental data. Hence, in light of the data available to date, coherent geosynchrotron emission is able to explain the bulk of the radio emission from EAS.

Our calculations show that LOPES should be able to easily detect the radio emission from a typical 10$^{17}$~eV air shower and will be a very useful tool for the study of EAS properties, especially the longitudinal structure of the particle distribution in the shower.

In the future, we plan to use this model as a basis for the development of a sophisticated numerical computation including, among other aspects,
\begin{itemize}
\item{near-field effects (important for ultra-high energy EAS that develop their maximum near ground level)}
\item{vectorial integration for analysis of the polarisation of the radiation}
\item{Askaryan-type \v Cerenkov emission}
\item{and an interface to realistic air shower simulations such as CORSIKA.}
\end{itemize}

\begin{acknowledgements}
We would like to thank Klaus Werner, Elmar K\"ording, Andreas Horneffer and Ralph Engel for a number of helpful discussions. We are also grateful for the comments of the referee that significantly helped us improve our paper. LOPES is supported by the German Federal Ministry of Education and Research under grant No.\ 05 CS1ERA/1 (Verbundforschung Astroteilchenphysik).
\end{acknowledgements}

\bibliography{h4423}
\bibliographystyle{aa}

\section*{Appendix: Geometry}

We adopt the instantaneous velocity vectors of the generated particle pairs as radially pointing away from the centre of the sphere. For a particle at position $(\vartheta,\varphi)$ on the shell, its direction is therefore given by
\begin{equation}
\vec{\hat{v}}(\vartheta,\varphi)= \left(\begin{array}{c}
\cos \eta \sin \vartheta \cos \varphi + \sin \eta \cos \vartheta \\
\sin \theta \sin \varphi \\
\sin \eta \sin \vartheta \cos \varphi - \cos \eta \cos \vartheta
\end{array}\right),
\end{equation}
whereas the direction of the $B$-field is given by
\begin{equation}
\vec{\hat{B}}= \left(\begin{array}{c}
\sin \eta_{B} \cos \varphi_{B} \\
\sin \eta_{B} \sin \varphi_{B} \\
- \cos \eta_{B}
\end{array}\right).
\end{equation}
Furthermore, the line-of-sight vector $\vec{R}$ from the particle to the observer is given by
\begin{equation}
\vec{R}(\vartheta,\varphi) = (R_{0}+K) \left(
\begin{array}{c}
\sin (\eta + \vartheta_{0}) \\
0 \\
-\cos (\eta + \vartheta_{0})
\end{array}
\right) - K \vec{\hat{v}}(\vartheta,\varphi)
\end{equation}
The direction of $\vec{R}$ is then calculated as
\begin{equation}
\vec{\hat{n}}(\vartheta,\varphi) = \frac{\vec{R}(\vartheta,\varphi)}{|\vec{R}(\vartheta,\varphi)|}
\end{equation}
and the pitch angle and angle to the line-of-sight correspond to
\begin{eqnarray}
\cos \alpha(\vartheta,\varphi) &=& \vec{\hat{v}}(\vartheta,\varphi) \cdot \vec{\hat{B}}\\
\cos \theta(\vartheta,\varphi) &=& \vec{\hat{v}}(\vartheta,\varphi) \cdot \vec{\hat{n}}(\vartheta,\varphi).\label{eqn:costheta}
\end{eqnarray}
The direction of the dominating emission component then changes as follows with $(\vartheta,\varphi)$:
\begin{equation}\label{eqn:epardir}
\vec{\hat{\mathrm{e}}}_{\parallel}(\vartheta,\varphi) = \frac{\vec{\hat{B}} \times \vec{\hat{v}}(\vartheta,\varphi)}{|\sin \alpha(\vartheta,\varphi)|}.
\end{equation}
These are all of the geometrical relations that are needed to execute the integration.

\end{document}